\pdfoutput=1  
\def\kl{\boldsymbol{\hat{J}}\boldsymbol\cdot\boldsymbol{\hat{l}}}
\def\km{\boldsymbol{\hat{J}}\boldsymbol\cdot\boldsymbol{\hat{m}}}
\def\kh{\boldsymbol{\hat{J}}\boldsymbol\cdot\boldsymbol{\hat{h}}}

\def\nk{n_\mathrm{K}}
\def\acap{\\ \nonumber \\}

\def\Pb{T_\mathrm{K}}
\def\rfr#1{Equation\,(\ref{#1})}
\def\rfrs#1#2{Equations\,(\ref{#1})--(\ref{#2})}
\def\Rfr#1{Equation\,(\ref{#1})}
\def\Rfrs#1#2{Equations\,(\ref{#1})--(\ref{#2})}
\def\derp#1#2{\rp{\partial{#1}}{\partial{#2}}}
\def\dert#1#2{\frac{{{\textrm{d}}}{#1}}{{{\textrm{d}}}{#2}}}
\def\virg#1{``#1"}
\def\eqi{\begin{equation}}
\def\eqf{\end{equation}}
\def\rp#1#2{\frac{#1}{#2}}
\def\lb#1{\label{#1}}

\def\ton#1{\left(#1\right)}
\def\qua#1{\left[#1\right]}
\def\grf#1{\left\{#1\right\}}
\def\ang#1{\left\langle #1\right\rangle}

\RequirePackage[2020-02-02]{latexrelease}
\documentclass{aastex631}
\usepackage{morefloats}
\usepackage[title]{appendix}
\usepackage{textcomp}
\usepackage{booktabs}
\usepackage{multirow}
\usepackage{rotating,tabularx}
\usepackage{float}
\usepackage{enumerate}
\usepackage{rotating}
\usepackage[polutonikogreek,english]{babel}
\usepackage{amsmath,starfont,textgreek,w-greek}
\usepackage[flushleft]{threeparttable}
\usepackage{amsthm}
\usepackage{bigints}
\usepackage{amscd}
\usepackage[mathlines]{lineno}
\usepackage{amssymb,dsfont}
\usepackage{graphicx,epsfig}
\usepackage{txfonts}
\usepackage{xr-hyper}
\usepackage{hyperref}

\usepackage[nointegrals]{wasysym}
\usepackage[caption=false]{subfig}

\allowdisplaybreaks

\makeatletter
 \DeclareRobustCommand\ref{%
    \@ifstar\@refstar\T@ref
  }%
  \DeclareRobustCommand\pageref{%
    \@ifstar\@pagerefstar\T@pageref
  }%
 \makeatother

\begin{document}

\title{When the anomalistic, draconitic and sidereal orbital periods do not coincide: the impact of post--Keplerian perturbing accelerations}

\shortauthors{L. Iorio}

\author[0000-0003-4949-2694]{Lorenzo Iorio}
\affiliation{Ministero dell' Istruzione e del Merito. Viale Unit\`{a} di Italia 68, I-70125, Bari (BA),
Italy}

\email{lorenzo.iorio@libero.it}

\begin{abstract}
In a purely Keplerian picture, the anomalistic, draconitic and sidereal orbital periods of a test particle orbiting a massive body coincide with each other. Such a degeneracy is removed when a post--Keplerian perturbing acceleration enters the equations of motion yielding generally different corrections to the Keplerian period for the three aforementioned characteristic orbital timescales. They are analytically worked out in the case of the accelerations induced by the general relativistic post--Newtonian gravitoelectromagnetic fields and, to the Newtonian level, by the oblateness of the central body as well. The resulting expressions hold for completely general orbital configurations and spatial orientations of the spin axis of the primary. Astronomical systems characterized by extremely accurate measurements of the orbital periods like, e.g., transiting exoplanets \textcolor{black}{and binary pulsars}, may offer potentially viable scenarios for measuring such post--Keplerian features of motion\textcolor{black}{, at least in principle}. As an example, the sidereal period of the \textcolor{black}{brown dwarf} WD1032 + 011 b is currently known with an uncertainty as small as $\simeq 10^{-5}\,\mathrm{s}$, while its predicted post--Newtonian gravitoelectric correction amounts to $0.07\,\mathrm{s}$\textcolor{black}{; however, the accuracy with which the Keplerian period can be calculated is just 572 s}. \textcolor{black}{For the double pulsar PSR J0737--3039, the largest relativistic correction to the anomalistic period amounts to a few tenths of a second, given a measurement error of such a characteristic orbital timescale as small as $\simeq 10^{-6}\,\mathrm{s}$. On the other hand, the Keplerian term can be currently calculated just to a $\simeq 9$ s accuracy. In principle, measuring  at least two of the three characteristic orbital periods for the same system independently would allow to cancel out their common Keplerian component provided that their difference is taken.}
\end{abstract}

\section{Introduction}
From a theoretical point of view, various time intervals $T$ characterizing different cyclic patterns of the orbital motion of a two--body gravitationally bound system can be defined when, in addition to the dominant Newtonian inverse--square acceleration,  also a much smaller, post--Keplerian\footnote{Here, by \virg{post--Keplerian}  one means dynamical features arising from any acceleration, Newtonian or not, different from the simple Newtonian inverse--square one. Then, in this specific sense, the classical acceleration due to, say, the primary's oblateness $J_2$ has to be meant as pK.} (pK) one $\boldsymbol{A}$ acts on a satellite. Such characteristic orbital timescales are the amounts of time elapsed between two successive passages of the latter at some directions which, in a purely Keplerian scenario, are all fixed; in this case, all such periods coincide with the Keplerian one $\Pb$. Instead, a pK perturbation breaks such a degeneracy, and the aforementioned temporal intervals generally differ one from each other.

Aim of the present work is to analytically work out the corrections $\Delta T$ to $\Pb$ induced by the gravitoelectromagnetic accelerations arising within the General Theory of Relativity (GTR) to the first post--Newtonian (1pN) order \citep{2001rfg..conf..121M,Mash07}. Furthermore, also the impact of the oblateness of the central body \citep{2005som..book.....C} is worked out to the Newtonian level. In all the aforementioned three cases, the anomalistic, draconitic and sidereal periods are considered. The resulting expressions turn out to be valid for completely general orbital shapes and inclinations, and for  arbitrary orientations of the primary's spin axis in space. For analogous calculation restricted to some particular orbital geometries, see \citet{2016MNRAS.460.2445I}. \textcolor{black}{In addition to the traditional quantities usually adopted like, e.g., the time-honoured pericentre precession, the orbital periods, if directly measured, may offer, in principle, further ways to test GTR and other models of gravity as well.}

\textcolor{black}{In recent years, exoplanets \citep{2008exop.book.....M,2010exop.book.....S,2012exop.book.....K,2018haex.bookE....D,2018exha.book.....P} have been attracting a growing interest as possible tools to test GTR and modified models of gravity
\citep{2006IJMPD..15.2133A,2006ApJ...649..992A,2006ApJ...649.1004A,2006NewA...11..490I,2008ApJ...685..543J,
2008MNRAS.389..191P,2009IAUS..253..492J,2009ApJ...698.1778R,2010OAJ.....3..167I,
2011A&A...535A.116D,2011PASJ...63..287F,2011Ap&SS.331..485I,
2011MNRAS.411..167I,2012MNRAS.423.1381E,2012ApJ...757..105K,2012Ap&SS.341..323L,
2013RAA....13.1231Z,2014MNRAS.438.1832X,2017PhLB..769..485V,2019A&A...628A..80B,
2019A&A...625A.121M,2020JCAP...06..042R,2021MNRAS.505.1567A,2021E&ES..658a2051G,2021PhRvD.104h4097K}}. \textcolor{black}{For comprehensive overviews of tests of GTR, see, e.g., \citet{2014LRR....17....4W,2018tegp.book.....W,Will20}, and references therein.} The results presented here may have an impact on\textcolor{black}{, e.g., just} exoplanetary studies since the accuracy in measuring the orbital periods of some transiting planets is currently quite remarkable. \textcolor{black}{Even better is the accuracy with which the anomalistic periods of binary pulsars are usually measured.}

The paper is organized as follows. In Section~\ref{Accel_sect}, the pK accelerations treated are reviewed: the 1pN gravitoelectric (Section~\ref{GE_sect}) and gravitomagnetic (Section~\ref{LT_sect}), and the Newtonian quadrupolar one (Section~\ref{J2_sect}). Section~\ref{Ano_sect} is devoted to the anomalistic period; the general calculational scheme is outlined in Section~\ref{ano_calc}, while the 1pN and the quadrupolar corrections are worked out in Sections~\ref{ano_GE} to \ref{ano_J2}.  The draconitic period is dealt with in Section~\ref{Dra_sect}; Section~\ref{dra_calc} shows how to calculate it, while the 1pN and the quadrupolar corrections are the subjects of Sections~\ref{dra_GE} to \ref{dra_J2}. The sidereal period is investigated in Section~\ref{Sid_sect}; the calculational approach is explained in Section~\ref{sid_calc}, while the 1pN and the quadrupolar corrections are calculated in Sections~\ref{sid_GE} to \ref{sid_J2}. In Section~\ref{num_sect}, a numerical evaluation of the 1pN gravitoelectric effect for a transiting exoplanet \textcolor{black}{and the double pulsar} whose orbital period\textcolor{black}{s are} accurately measured is offered. Section~\ref{fine} summarizes the findings  and offers conclusions.
\section{The pK accelerations}\lb{Accel_sect}
Here, a brief summary of some key concepts of celestial mechanics, needed to follow the rest of the paper profitably, is offered \citep{1961mcm..book.....B,Sof89,1991ercm.book.....B,2003ASSL..293.....B,2005ormo.book.....R,2011rcms.book.....K,2014grav.book.....P,SoffelHan19}.

If $\boldsymbol{A}$ is an arbitrary pK perturbing acceleration, generally depending on the position and velocity vectors $\boldsymbol{r}$ and $\boldsymbol{v}$ of the orbiter,  calculating its effects on the orbital path of the  latter requires the knowledge of its radial, transverse and normal components $A_r,\,A_\tau$ and $A_h$. They are the projections $\boldsymbol{A}\boldsymbol{\cdot}\boldsymbol{\hat{r}},\,\boldsymbol{A}\boldsymbol{\cdot}\boldsymbol{\hat{\tau}}$ and $\boldsymbol{A}\boldsymbol{\cdot}\boldsymbol{\hat{h}}$ of $\boldsymbol{A}$ onto the co--moving radial, transverse and normal unit vectors
\begin{align}
\boldsymbol{\hat{r}} \lb{ur} & = \grf{\cos\Omega\cos u - \cos I\sin\Omega\sin u,\,\sin\Omega\cos u + \cos I\cos\Omega\sin u,\,\sin I\sin u},\acap
\boldsymbol{\hat{\tau}} \lb{ut} & = \grf{ -\cos\Omega\sin u - \cos I\sin\Omega\cos u,\,-\sin\Omega\sin u + \cos I\cos\Omega\cos u,\,\sin I\,\cos u},\acap
\boldsymbol{\hat{h}} \lb{un} & = \grf{\sin I\sin\Omega,\,-\sin I\cos\Omega,\,\cos I}.
\end{align}
In \rfrs{ur}{un}, $I$ is the inclination of the orbital plane to the reference plane of the coordinate system adopted, $\Omega$ is the longitude of the ascending node\footnote{It is an angle counted in the reference plane from the reference $x$ direction to the line of nodes, which is the intersection of the orbital plane with the fundamental one.}, and $u$ is the argument of latitude\footnote{It is a time--dependent angle reckoned in the orbital plane from the line of nodes to the instantaneous position of the test particle moving along the ellipse.} defined as
\eqi
u:=\omega + f.\lb{arglat}
\eqf
In \rfr{arglat}, $\omega$ is the argument of pericentre\footnote{It is an angle reckoned in the orbital plane from the line of nodes to the position of the pericentre on the line of apsides.}, and $f$ is the true anomaly\footnote{It is a time--dependent angle counted in the orbital plane from the position of the pericentre on the line of apsides to the instantaneous position of the test particle moving along the ellipse. \textcolor{black}{Thus, it is often used as fast variable of integration when the average over one orbital period of some relevant quantity instantaneously varying during the satellite's orbital revolution is calculated.}}. The normal unit vector, given by \rfr{un}, is aligned with the orbital angular momentum; \rfrs{ur}{un} are connected at each instant of time by the relation
\eqi
\boldsymbol{\hat{h}} \boldsymbol{\times}\boldsymbol{\hat{r}}= \boldsymbol{\hat{\tau}}.
\eqf
Then,  $A_r,\,A_\tau$ and $A_h$ are to be evaluated onto the Keplerian ellipse, assumed as unperturbed reference trajectory; it is given by
\eqi
r = \rp{p}{1+ e\cos f}.\lb{rKep}
\eqf
In \rfr{rKep}, $e$ is the eccentricity\footnote{It fixes the shape of the ellipse; $0\leq e<1$, where $e=0$ corresponds to a circular orbit.}, and
\eqi
p:=a\ton{1-e^2}
\eqf
is the orbit's semilatus rectum; $a$ is the semimajor axis\footnote{It determines the size of the ellipse}.
The position and velocity vectors, generally entering the analytical expression of $\boldsymbol{A}$, can be conveniently expressed as
\begin{align}
\boldsymbol{r} \lb{rvec} & = r\ton{\boldsymbol{\hat{l}}\cos u + \boldsymbol{\hat{m}}\sin u}, \acap
\boldsymbol{v} \lb{vvec}& = \sqrt{\rp{\upmu}{p}}\qua{-\boldsymbol{\hat{l}}\ton{e \sin\omega + \sin u}  + \boldsymbol{\hat{m}}\ton{e \cos\omega + \cos u} }.
\end{align}
In \rfrs{rvec}{vvec}, $r$ is given by \rfr{rKep}, while
\begin{align}
\boldsymbol{\hat{l}} \lb{elle}& := \grf{\cos\Omega,~\sin\Omega,~0},\acap
\boldsymbol{\hat{m}} \lb{emme}& := \grf{-\cos I\sin\Omega,~\cos I\cos\Omega,~\sin I}
\end{align}
are two unit vectors lying in the orbital plane; $\boldsymbol{\hat{l}}$ is directed along the line of nodes, while $\boldsymbol{\hat{m}}$ is perpendicular to $\boldsymbol{\hat{l}}$ in such a way that
\eqi
\boldsymbol{\hat{l}}\boldsymbol{\times}\boldsymbol{\hat{m}} = \boldsymbol{\hat{h}}.
\eqf
Finally,
\eqi
\upmu:=GM
\eqf
entering \rfr{vvec} is the standard gravitational parameter of the primary; $M$ is its mass, and $G$ is the Newtonian constant of gravitation.
\subsection{The 1pN gravitoelectric acceleration}\lb{GE_sect}
In the case of a binary system made of two non--rotating bodies A and B of finite masses $M_\mathrm{A}$ and $M_\mathrm{B}$, the 1pN gravitoelectric acceleration is
\citetext{\citealp[see, e.g.,][Equation\,(2.5),\,p.\,111]{1985AIHPA..43..107D}; \citealp[Equation\,(A2.6),\,p.\,166]{Sof89}; \citealp[Equation\,(4.4.28),\,p.\,154]{1991ercm.book.....B}; \citealp[Equation\,(10.3.7),\,p.\,381]{SoffelHan19}; \citealp[Equation\,(10.1),\,p.\,482]{2014grav.book.....P}}
\begin{align}
{\boldsymbol{A}}^\mathrm{1pN} \lb{A1pN} &= \rp{\upmu_\mathrm{b}}{c^2 r^2}\grf{
\qua{ \ton{4 + 2\nu}\rp{\upmu_\mathrm{b}}{r} + \rp{3}{2}\nu v_r^2 - \ton{1 - 3\nu}v^2 }\boldsymbol{\hat{r}} + \ton{4 - 2\nu}v_r\boldsymbol{v}
}.
\end{align}
In \rfr{A1pN}, $c$ is the speed of light in vacuum, and
\eqi
\upmu_\mathrm{b}:=GM_\mathrm{b}
\eqf
is the standard gravitational parameter of the binary whose total mass is
\eqi
M_\mathrm{b}:= M_\mathrm{A} + M_\mathrm{B}.
\eqf
Moreover,
\eqi
\nu: = \rp{M_\mathrm{A}M_\mathrm{B}}{M^2_\mathrm{b}}
\eqf
is the binary's symmetric mass ratio ranging from $0$ if one of the two bodies can be considered as a test particle to $1/4\textcolor{black}{=0.25}$ if both bodies have the same mass, $\boldsymbol{r}$ is the position vector of one body with respect to the other one, $\boldsymbol{\hat{r}}$ is its unit vector, $r$ is its magnitude giving the relative distance between the two bodies, $\boldsymbol{v}$ is the relative velocity between them, whose magnitude is $v$, while \eqi
v_r:=\boldsymbol{v}\boldsymbol{\cdot}\boldsymbol{\hat{r}}
\eqf
is the projection of the relative velocity  onto the position unit vector.
By using \rfrs{ur}{arglat} and \rfrs{rKep}{emme}, the radial, transverse and normal components $A_r^\mathrm{1pN},\,A_{\tau}^\mathrm{1pN},\,A_h^\mathrm{1pN}$ of \rfr{A1pN} turn out to be
\begin{align}
A_r^\mathrm{1pN} \lb{Ar1pN} & = \rp{\upmu^2_\mathrm{b}\ton{1 + e \cos f}^2}{4 c^2 a^3\ton{1 - e^2}^3}\qua{
e^2 \ton{4 - 13 \nu} - 4 \ton{- 3 + \nu}  + 8 e \ton{1 - 2 \nu} \cos f + e^2 \ton{- 8 + \nu} \cos 2 f
}, \acap
A_\tau^\mathrm{1pN} \lb{At1pN}& = \rp{2 e \upmu^2_\mathrm{b}\ton{1 + e \cos f}^3 \ton{2 - \nu}\sin f}{c^2 a^3\ton{1 - e^2}^3}, \acap
A_h^\mathrm{1pN} \lb{Ah1pN}& = 0.
\end{align}
\Rfrs{Ar1pN}{Ah1pN} agree with, e.g., Equations (A2.77a)--(A2.77c), calculated with GTR, by \citet[p.\,178]{Sof89}.
\subsection{The 1pN gravitomagnetic Lense--Thirring acceleration}\lb{LT_sect}
The 1pN gravitomagnetic Lense--Thirring (LT) acceleration due to the angular momentum $\boldsymbol{J}$ of a massive primary, is, for an arbitrary orientation of the latter, \citep{Sof89,1990CeMDA..48..167H,1994PhRvD..49..618D,iers10,2014grav.book.....P,SoffelHan19}
\eqi
{\boldsymbol{A}}^\mathrm{LT} = \rp{2GJ}{c^2 r^3}\qua{3\ton{\boldsymbol{\hat{J}}\boldsymbol{\cdot}\boldsymbol{\hat{r}}}\boldsymbol{\hat{r}}\boldsymbol{\times}\boldsymbol{v} + \boldsymbol{v}\boldsymbol{\times}\boldsymbol{\hat{J}}}\lb{ALT}.
\eqf
In \rfr{ALT}, $\boldsymbol{\hat{J}}$ is the primary's spin axis which can be parameterized as, e.g.,
\begin{align}
{\hat{J}}_x \lb{Jx}& = \cos\alpha_J\cos\delta_J,\ \acap
{\hat{J}}_y \lb{Jy}& = \sin\alpha_J\cos\delta_J,\ \acap
{\hat{J}}_z \lb{Jz}& = \sin\delta_J,
\end{align}
where $\alpha_J$ and $\delta_J$ are its longitude and latitude angles in some coordinate system.
For a generalization of \rfr{ALT} to a two--body system with comparable masses and spins, see, e.g., \citet[Equation\,(2.2.c)]{1995PhRvD..52..821K}, and \citet{Sof89}. Basically, if ${\boldsymbol{J}}_\mathrm{A}$ and ${\boldsymbol{J}}_\mathrm{B}$ are the spin angular momenta of the extended bodies A and B, $\boldsymbol{J}$ has to be replaced in \rfr{ALT} and in all the following Equations with
\eqi
{\boldsymbol{S}}_\mathrm{b}:= \ton{1 + \rp{3}{4}\rp{M_\mathrm{B}}{M_\mathrm{A}}}{\boldsymbol{J}}_\mathrm{A} +  \ton{1 + \rp{3}{4}\rp{M_\mathrm{A}}{M_\mathrm{B}}}{\boldsymbol{J}}_\mathrm{B}
\eqf

It is useful to define the following quantities:
\begin{align}
\texttt{Jl}\lb{Kal} := \kl, \acap
\texttt{Jm}\lb{Kam} := \km, \acap
\texttt{Jh}\lb{Kah} := \kh.
\end{align}
By means of \rfrs{ur}{arglat}, \rfrs{rKep}{vvec} and \rfrs{Kal}{Kah}, the radial, transverse and normal components $A_r^\mathrm{LT},\,A_{\tau}^\mathrm{LT},\,A_h^\mathrm{LT}$ of \rfr{ALT} can be written as
\begin{align}
A_r^\mathrm{LT} \lb{ARLT} & = \rp{2\nk G J\ton{1 + e\cos f}^4\texttt{Jh}}{c^2 a^2\ton{1 - e^2}^{7/2}}, \\ \nonumber \\
A_{\tau}^\mathrm{LT} \lb{ATLT} & = -\rp{2e\nk G J\ton{1 + e\cos f}^3\sin f\texttt{Jh}}{c^2 a^2\ton{1 - e^2}^{7/2}}, \\ \nonumber \\
A_h^\mathrm{LT}  \lb{ANLT} & = -\rp{2\nk G J\ton{1 + e\cos f}^3}{c^2 a^2\ton{1 - e^2}^{7/2}}\grf{
\qua{e\cos\omega - \ton{2 + 3 e \cos f} \cos u}\texttt{Jl} - \rp{1}{2}\qua{e\sin\omega + 4\sin u + 3 e \sin\ton{2f + \omega}}\texttt{Jm}
},
\end{align}
where
\eqi
\nk:=\sqrt{\rp{\upmu}{a^3}} = \rp{2\uppi}{\Pb}
\eqf
is the Keplerian mean motion.
\subsection{The Newtonian quadrupolar acceleration}\lb{J2_sect}
The pK acceleration induced by the first even zonal harmonic coefficient $J_2$ of the multipolar expansion of the exterior Newtonian gravitational potential of a massive primary endowed with axial symmetry is
\eqi
{\boldsymbol{A}}^{J_2} = \rp{3\upmu J_2 R^2_\mathrm{e}}{2r^4}\grf{\qua{5\ton{\boldsymbol{\hat{J}}\boldsymbol{\cdot}\boldsymbol{\hat{r}}}^2 -1}\boldsymbol{\hat{r}} - 2\ton{\boldsymbol{\hat{J}}\boldsymbol{\cdot}\boldsymbol{\hat{r}}}\boldsymbol{\hat{J}}},\lb{AJ2}
\eqf
where $R_\mathrm{e}$ is the body's equatorial radius.

By defining
\begin{align}
\widehat{T}_1 \lb{T1}&:= 1, \\ \nonumber \\
\widehat{T}_2 \lb{T2}&:= \texttt{Jl}^2 + \texttt{Jm}^2,\\ \nonumber \\
\widehat{T}_3 \lb{T3}&:= \texttt{Jl}^2 - \texttt{Jm}^2,\\ \nonumber \\
\widehat{T}_4 \lb{T4}&:= \texttt{Jh}\,\texttt{Jl},\\ \nonumber \\
\widehat{T}_5 \lb{T5}&:= \texttt{Jh}\,\texttt{Jm},\\ \nonumber \\
\widehat{T}_6 \lb{T6}&:= \texttt{Jl}\,\texttt{Jm},
\end{align}
and by means of \rfrs{ur}{arglat} and \rfrs{rKep}{rvec}, the radial, transverse and normal components $A_r^{J_2},\,A_{\tau}^{J_2},\,A_h^{J_2}$ of \rfr{AJ2} can be cast into the form
\begin{align}
A_r^{J_2} \lb{ARJ2} & = \rp{3\upmu J_2 R^2_\mathrm{e}\ton{1 + e\cos f}^4}{2 a^4\ton{1 - e^2}^4}\qua{- \widehat{T}_1  +  3\ton{\rp{\widehat{T}_2}{2}  + \rp{\widehat{T}_3 \cos 2u}{2} + \widehat{T}_6\sin 2u}}, \\ \nonumber \\
A_{\tau}^{J_2} \lb{ATJ2} & = \rp{3\upmu J_2 R^2_\mathrm{e}\ton{1 + e\cos f}^4}{a^4\ton{1 - e^2}^4}\ton{\rp{\widehat{T}_3\sin 2u}{2} - \widehat{T}_6\cos 2u}, \\ \nonumber \\
A_{h}^{J_2} \lb{ANJ2} & = -\rp{3\upmu J_2 R^2_\mathrm{e}\ton{1 + e\cos f}^4}{a^4\ton{1 - e^2}^4}\ton{\widehat{T}_4 \cos u + \widehat{T}_5 \sin u}.
\end{align}

If both bodies A and B are extended and axisymmetric, \rfr{AJ2} becomes \citep{1975PhRvD..12..329B}
\eqi
{\boldsymbol{A}}^{J_2} = \rp{3\upmu_\mathrm{b}}{2r^4}\mathcal{F}_\mathrm{AB},\lb{AQ2}
\eqf
where
\eqi
\mathcal{F}_\mathrm{AB}:= J_2^\mathrm{A} {R^\mathrm{A}_\mathrm{e}}^2\grf{\qua{5\ton{{\boldsymbol{\hat{J}}}^\mathrm{A}\boldsymbol{\cdot}\boldsymbol{\hat{r}}}^2 -1}\boldsymbol{\hat{r}} - 2\ton{{\boldsymbol{\hat{J}}}^\mathrm{A}\boldsymbol{\cdot}\boldsymbol{\hat{r}}}{\boldsymbol{\hat{J}}}^\mathrm{A}} + J_2^\mathrm{B} {R^\mathrm{B}_\mathrm{e}}^2\grf{\qua{5\ton{{\boldsymbol{\hat{J}}}^\mathrm{B}\boldsymbol{\cdot}\boldsymbol{\hat{r}}}^2 -1}\boldsymbol{\hat{r}} - 2\ton{{\boldsymbol{\hat{J}}}^\mathrm{B}\boldsymbol{\cdot}\boldsymbol{\hat{r}}}{\boldsymbol{\hat{J}}}^\mathrm{B}},\lb{FAB}
\eqf
and $r$ and $\boldsymbol{\hat{r}}$ entering \rfrs{AQ2}{FAB} refer to the relative orbit.

\section{The apsidal period}\lb{Ano_sect}
\subsection{General calculational scheme}\lb{ano_calc}
The anomalistic period $T_\mathrm{ano}$ is defined as the time interval between
two successive instants when the real position of the test particle coincides with the pericentre position on the corresponding orbit.
It can be calculated as \citep{Zhongo60,1979AN....300..313M,2016MNRAS.460.2445I}
\eqi
T_\mathrm{ano} =\Pb + \Delta T_\mathrm{ano} = \bigintss_0^{2\uppi}\ton{\dert{t}{f}}df,\lb{mioc}
\eqf
where $dt/df$, when a pK acceleration $\boldsymbol{A}$ is present,  is given by
\eqi
\dert t f \simeq \rp{r^2}{\sqrt{\upmu p}}\qua{1 + \rp{r^2}{\sqrt{\upmu p}}\ton{\dert\omega t + \cos I\dert\Omega t}},\lb{dtdf2}
\eqf
since \citep{1958SvA.....2..147E,1959ForPh...7S..55T,1979AN....300..313M,1991ercm.book.....B,2003ASSL..293.....B,2014grav.book.....P}
\eqi
\dert f t = \rp{\sqrt{\upmu p}}{r^2}\qua{1 - \rp{r^2}{\sqrt{\upmu p}}\ton{\dert\omega t + \cos I\dert\Omega t}}.\lb{dfdt}
\eqf
The true anomaly $f$ enters \rfr{mioc} as fast variable of integration just because the line of apsides is involved in the definition of anomalistic period.
In order to obtain the full correction $\Delta T_\mathrm{ano}$ of the order of $A$ to the Keplerian  orbital period, the contribution of the second term of \rfr{dtdf2} to \rfr{mioc} is not enough. Indeed, also
the partial derivatives of the Keplerian term of \rfr{dtdf2}
\eqi
\left.\dert t f\right|_\mathrm{K} = \rp{r^2}{\sqrt{\upmu p}}\lb{dtdf}
\eqf
with respect to $a$ and $e$, multiplied by the finite variations $\Delta a\ton{f},\,\Delta e\ton{f}$  of the same orbital elements, have to be taken; in this way, one fully accounts for the fact that the Keplerian  orbital elements vary instantaneously as the satellite goes along its trajectory.
Thus, it is finally obtained
\begin{align}
\Delta T_\mathrm{ano} \lb{loro} &= \bigint_0^{2\uppi}\grf{ \rp{3}{2}\sqrt{\rp{a\ton{1 - e^2}^3}{\upmu}}\rp{\Delta a\ton{f}}{\ton{1+e\cos f}^2} - \sqrt{\rp{a^3\ton{1-e^2}}{\upmu}}\rp{\qua{3 e + \ton{2 + e^2}\cos f}}{\ton{1 + e\cos f}^3}\Delta e\ton{f} + \rp{r^4}{\upmu p}\ton{\dert\omega t  + \cos I\dert\Omega t}}_\mathrm{K}df.
\end{align}
The suffix K in \rfr{loro} means that the content of the curly brackets to which it is appended has to be evaluated onto the unperturbed Keplerian ellipse given by \rfr{rKep}.
Furthermore, the instantaneous shifts $\Delta\upkappa\ton{f}$ of $\upkappa=a,e$ are calculated, to the first order in the perturbing pK acceleration $A$, as
\eqi
\Delta\upkappa\ton{f}=\int_{f_0}^{f^{'}}\dert{\upkappa}{f^{'}}df^{'}=\int_{f_0}^{f^{'}}\dert{\upkappa}{t}\dert t{f^{'}}df^{'},
\eqf
where $d\upkappa/dt$ are the Gauss equations for the variations of $\upkappa=a,e$ \citep{1961mcm..book.....B,Sof89,1991ercm.book.....B,2003ASSL..293.....B,2005ormo.book.....R,2011rcms.book.....K,2014grav.book.....P,SoffelHan19}
\begin{align}
\dert{a}{t} \lb{dadt} &= \rp{2}{\nk \sqrt{1-e^2}} \qua{e A_{r} \sin f + \ton{\rp{p}{r}} A_{\tau}},\\ \nonumber \\
\dert e t \lb{dedt} & = \rp{\sqrt{1-e^2}}{\nk a} \grf{A_{r} \sin f + A_{\tau} \qua{\cos f + \rp{1}{e} \ton{1-\rp{r}{a}} }}.
\end{align}
Finally, $d\Omega/dt$ and $d\omega/dt$ entering the third term of \rfr{loro} are the Gauss equations for the variations of the longitude of the ascending node and the argument of pericentre, respectively, given by \citep{1961mcm..book.....B,Sof89,1991ercm.book.....B,2003ASSL..293.....B,2005ormo.book.....R,2011rcms.book.....K,2014grav.book.....P,SoffelHan19}
\begin{align}
\dert \Omega t \lb{dOdt}& = \rp{1}{\nk a \sin I \sqrt{1-e^2}} A_{h} \ton{\rp{r}{a}} \sin u, \\ \nonumber\\
\dert \omega t \lb{dodt} & = \rp{\sqrt{1-e^2}}{\nk a e} \qua{-A_{r} \cos f + A_{\tau} \ton{1 + \rp{r}{p}} \sin f} - \cos I \dert\Omega t.
\end{align}

In \citep{2016MNRAS.460.2445I}, a variant of the above calculation can be found; in \rfr{dtdf}, $p$ is adopted as independent variable along with the eccentricity $e$, and simpler expressions for the partial derivatives of \rfr{dtdf} are obtained. The resulting expressions for calculating $\Delta T_\mathrm{ano}$ turns out to be
\begin{align}
\Delta T_\mathrm{ano} \lb{mia} &= \bigint_0^{2\uppi}\grf{ \rp{3}{2}\sqrt{\rp{p}{\upmu}}\rp{\Delta p\ton{f}}{\ton{1+e\cos f}^2} - 2\sqrt{\rp{p^3}{\upmu}}\rp{\cos f\Delta e\ton{f}}{\ton{1 + e\cos f}^3} + \rp{r^4}{\upmu p}\ton{\dert\omega t  + \cos I\dert\Omega t}}_\mathrm{K}df.
\end{align}
The first--order variation $\Delta p\ton{f}$ of the semilatus rectum can be calculated from \citep{1959ForPh...7S..55T,1979AN....300..313M}
\eqi
\dert p f = \rp{2 r^3 A_{\tau}}{\upmu}.
\eqf
In the end, both \rfr{loro} and \rfr{mia} give the same result.

The presence or not of the pK anomalistic correction $\Delta T_\mathrm{ano}$ to the orbital period can be intuitively explained as follows.  According to
\eqi
\eta = \nk\ton{t_0 - t_\mathrm{p}},\lb{kazzo}
\eqf
where $\eta$ is the mean anomaly at epoch, $t_0$ is the initial epoch and $t_\mathrm{p}$ is the time of passage at pericentre, the rate of change of the $\eta$ is proportional to the opposite of the pace of variation of  $t_\mathrm{p}$. Thus, should $\eta$ increase, the crossing of the pericentre position would be anticipated with respect to the Keplerian  case since $t_\mathrm{p}$ would decrease, and vice versa. In this case, the variation of $\eta$ would result in an orbit--by--orbit advance or delay of the passages at the pericentre. As it will be shown, while the 1pN gravitoelectric acceleration of \rfr{A1pN} does induce a negative rate of $\eta$, the gravitomagnetic LT one of \rfr{ALT} leaves the mean anomaly at epoch unchanged. Furthermore, several modified models of gravity, inducing radial pK accelerations dependent only on $r$, secularly change both $\omega$ and $\eta$. Also the Newtonian acceleration raised by the primary's oblateness $J_2$ affects, among other things, also $\eta$.
\subsection{The 1pN gravitoelectric correction}\lb{ano_GE}
The 1pN anomalistic period can be calculated by means of \rfrs{Ar1pN}{Ah1pN} as explained in Section~\ref{ano_calc}. It turns out to be
\eqi
T_\mathrm{ano}^\mathrm{1pN} = \Pb + \Delta T_\mathrm{ano}^\mathrm{1pN},\lb{Panol}
\eqf
with
\begin{align}
\Delta T_\mathrm{ano}^\mathrm{1pN} \nonumber & = \rp{\uppi\sqrt{\upmu_\mathrm{b} a}}{2 c^2\ton{1 - e^2}^2}\ton{
36 + e^2\ton{42 - 38 \nu} + 2 e^4\ton{6 - 7 \nu} - 8 \nu + 3 e\grf{\qua{28 + 3 e^2\ton{4 - 5 \nu} - 12 \nu} \cos f_0 - \right.\right.\acap
\lb{anoGENU} &\left.\left. - e\ton{-10 + 8 \nu + e \nu \cos f_0 }\cos 2 f_0 }}.
\end{align}
In \rfr{anoGENU}, $f_0$ is the true anomaly at the initial epoch.
In the test particle limit $\nu\rightarrow 0$, \rfr{anoGENU} reduces to
\eqi
\Delta T^\mathrm{1pN}_\mathrm{ano} \lb{TanoGE} = \rp{3\,\pi\,\sqrt{\upmu a}}{c^2\,\ton{1 - e^2}^2}\,\qua{6 + 7\,e^2 + 2\,e^4 + 2\,e\,\ton{7 + 3\,e^2}\,\cos f_0 + 5\,e^2\,\cos 2f_0}.
\eqf

\begin{figure}
\centering
\includegraphics[width=\columnwidth]{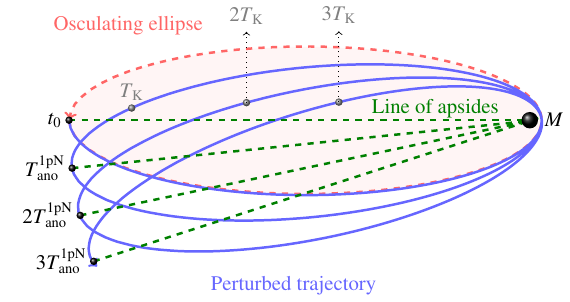}
\caption{Perturbed 1pN trajectory (continuous blue curve)  and its osculating
Keplerian ellipse (dashed red curve) at the initial instant of time $t_0$ of a restricted two--body system characterized by $e = 0.95,\,I = 0,\,\Omega = 0,\omega=90^\circ,\,f_0 = 180^\circ$ as seen from above the fixed orbital plane. Here, it is assumed that both $\omega$ and $\eta$ undergo their known 1pN gravitoelectric secular precessions  due to the mass $M$ of the primary \citep{2017EPJC...77..439I}. For a better visualization of their effect, their sizes are suitably rescaled.
The positions on the perturbed trajectory after one, two and three Keplerian periods $\Pb$ are marked in gray. At each orbit, the passages at the precessing dashed green line of apsides occur always later than in the Keplerian case by the amount given by \rfr{TanoGE}, which is always positive.
}\label{fig_per_GE_ano}
\end{figure}

Figure \ref{fig_per_GE_ano} is obtained for generic values of the Keplerian orbital elements of a test particle revolving around a massive primary. It confirms the analytical result of \rfr{TanoGE}; over, say, three orbital revolutions, the satellite reaches always the precessing line of apsides after a time interval equal to $T_\mathrm{ano}^\mathrm{1pN}$. It is longer than $\Pb$, in agreement with \rfr{TanoGE}, which is always positive.

Furthermore, Figure \ref{P_ano_GE} plots  the final part of the time series of the cosine $\boldsymbol{\hat{r}}\boldsymbol{\cdot}\boldsymbol{\hat{C}}$ of the angle between the position vector $\boldsymbol{r}$ and the
Laplace--Runge--Lenz unit vector $\boldsymbol{\hat{C}}$ versus time $t$, in units of $\Pb$, for a numerically integrated fictitious test particle with and without \rfr{A1pN} starting in both cases from, say, the moving pericentre, i.e., for ${\boldsymbol{\hat{r}}}_0\boldsymbol{\cdot}{\boldsymbol{\hat{C}}}_0 = +1$. It can be seen that the orbiter comes back to the same position on the precessing line of apsides, i.e. it is $\boldsymbol{\hat{r}}\boldsymbol{\cdot}\boldsymbol{\hat{C}} = +1$ again, just after   $T^\mathrm{1pN}_\mathrm{ano} = \Pb + \Delta T^\mathrm{1pN}_\mathrm{ano}$  differing from $\Pb$ by a (positive) amount, in agreement with \rfr{TanoGE}.
\begin{figure}
\centering
\includegraphics[width=\columnwidth]{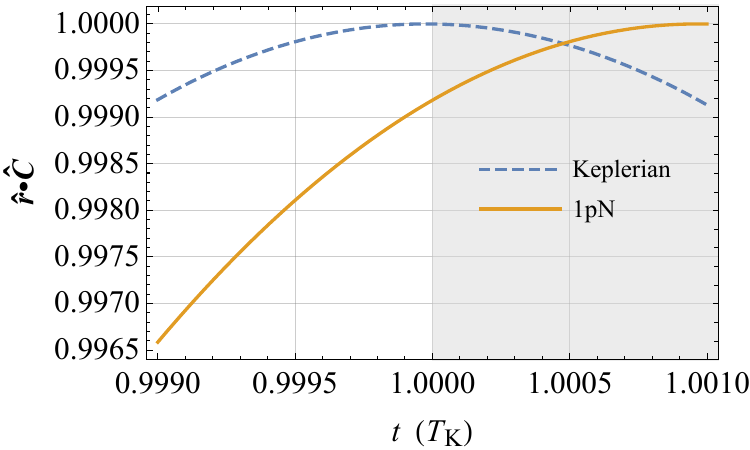}
\caption{Numerically produced time series of the cosine $\boldsymbol{\hat{r}}\boldsymbol{\cdot}\boldsymbol{\hat{C}}$ of the angle between the position vector $\boldsymbol{r}$ and the Laplace--Runge--Lenz vector $\boldsymbol{C}$  versus time $t$, in units of $\Pb$, obtained by integrating the equations of motion of a fictitious test particle with (continuous ocra yellow curve) and without (dashed azure curve) the 1pN gravitoelectric acceleration of \rfr{A1pN} for an elliptical ($e=0.665$)  orbit arbitrarily oriented in space ($I = 40^\circ,\,\Omega = 45^\circ,\,\omega = 50^\circ$) starting from the periapsis ($f_0 = 0$), i.e., ${\boldsymbol{\hat{r}}}_0\boldsymbol\cdot{\boldsymbol{\hat{C}}}_0 = +1$; the semimajor axis is $a = 6 R_\mathrm{e}$. The physical parameters of the Earth are adopted. The  1pN acceleration is suitably rescaled in such a way that $\Delta T^\mathrm{1pN}_\mathrm{ano}/\Pb=0.001$. The time needed to come back to the initial position on the (moving) line of apsides, so that $\boldsymbol{\hat{r}}\boldsymbol{\cdot}\boldsymbol{\hat{C}} = +1$ again, is longer than in the Keplerian case by the amount $\Delta T^\mathrm{1pN}_\mathrm{ano} = +0.001\Pb$, shown by the shaded area, in agreement with \rfr{TanoGE}. }\lb{P_ano_GE}
\end{figure}
\subsection{The 1pN gravitomagnetic Lense--Thirring correction}\lb{ano_LT}
The LT anomalistic period can be calculated by means of \rfrs{ARLT}{ANLT} as explained in Section~\ref{ano_calc}. It turns out to be
\eqi
T_\mathrm{ano}^\mathrm{LT} = \Pb + \Delta T_\mathrm{ano}^\mathrm{LT},
\eqf
with
\eqi
\Delta T^\mathrm{LT}_\mathrm{ano} = 0\lb{TanoLT};
\eqf
it is an exact result, valid to all orders in the eccentricity $e$.

\begin{figure}
\centering
\includegraphics[width=\columnwidth]{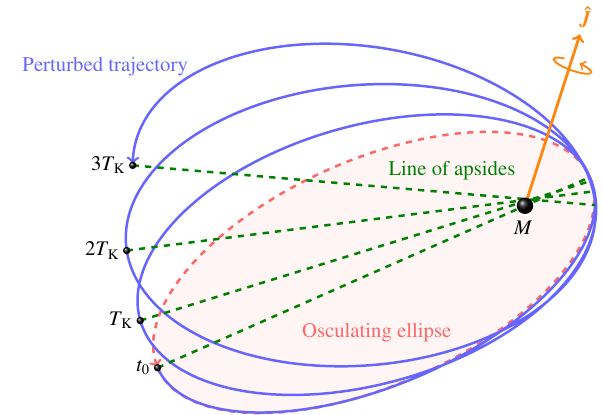}
\caption{Perturbed LT trajectory (continuous blue curve)  and its osculating
Keplerian ellipse (dashed red curve)  at the initial instant of time $t_0$ characterized by $e = 0.7,\,I = 30^\circ,\,\Omega = 72^\circ,\,\omega = 50^\circ,\,f_0 = 180^\circ$. The orientation of the spin axis $\boldsymbol{\hat{J}}$ of the central body is set by $\alpha_J=45^\circ,\,\delta_J=60^\circ$. In this example, $I$, $\Omega$ and $\omega$ undergo their known LT precessions due to the spin angular momentum $\boldsymbol{J}$ of the primary \citep{2017EPJC...77..439I}; their magnitudes are suitably rescaled by enhancing them for a better visualisation. The initial position is chosen at the apocentre instead of the pericentre solely for the sake of better visualisation.
The positions on the perturbed trajectory after one, two and three Keplerian periods are marked. At each orbit, the passage at the  drifting dashed green line of apsides occurs always as in the Keplerian case because, according to \rfr{TanoLT}, $\Delta T_\mathrm{ano}^\mathrm{LT}=0$.
}\label{fig_per_LT_ano}
\end{figure}

Figure \ref{fig_per_LT_ano}, obtained for generic values of the Keplerian orbital parameters, shows just that; over three orbital revolutions, the test particle reaches always the precessing line of apsides after a time interval equal to the Keplerian  orbital period after each orbit.

Furthermore, Figure \ref{P_ano_LT} plots  the final part of the time series of the cosine $\boldsymbol{\hat{r}}\boldsymbol{\cdot}\boldsymbol{\hat{C}}$ of the angle between the position vector $\boldsymbol{r}$ and the Laplace--Runge--Lenz vector $\boldsymbol{C}$ versus time $t$, in units of $\Pb$, for a numerically integrated fictitious test particle acted upon by \rfr{ALT} starting from, say, the moving pericentre, i.e., for ${\boldsymbol{\hat{r}}}_0\boldsymbol{\cdot}{\boldsymbol{\hat{C}}}_0 = +1$. It can be seen that it comes back to the same position on the precessing line of apsides, i.e. it is $\boldsymbol{\hat{r}}\boldsymbol{\cdot}\boldsymbol{\hat{C}} = +1$ again, just after one Keplerian orbital period.
\begin{figure}
\centering
\includegraphics[width=\columnwidth]{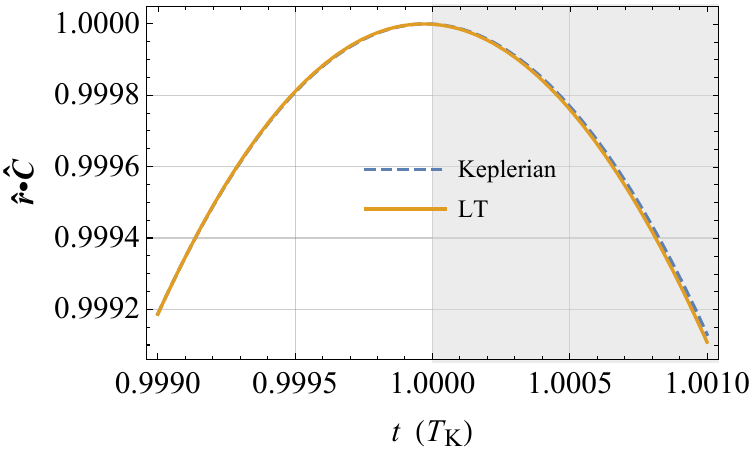}
\caption{Numerically produced time series of the cosine $\boldsymbol{\hat{r}}\boldsymbol{\cdot}\boldsymbol{\hat{C}}$ of the angle between the position vector $\boldsymbol{r}$ and the Laplace--Runge--Lenz vector $\boldsymbol{C}$  versus time $t$, in units of $\Pb$, obtained by integrating the equations of motion of a fictitious test particle with (continuous ocra yellow curve) and without (dashed azure curve) the LT acceleration of \rfr{ALT} for an elliptical ($e=0.665$)  orbit arbitrarily oriented in space ($I = 40^\circ,\,\Omega = 45^\circ,\,\omega = 50^\circ$) starting from the periapsis ($f_0 = 0$), i.e., ${\boldsymbol{\hat{r}}}_0\boldsymbol\cdot{\boldsymbol{\hat{C}}}_0 = +1$; the semimajor axis is $a = 6 R_\mathrm{e}$. The physical parameters of the Earth are adopted, apart from the spin axis position set by $\alpha_J=45^\circ,\,\delta_J=60^\circ$. The time needed to come back to the initial position on the (moving) line of apsides, so that $\boldsymbol{\hat{r}}\boldsymbol{\cdot}\boldsymbol{\hat{C}} = +1$ again, is as in the Keplerian case.}\lb{P_ano_LT}
\end{figure}

The fact that the gravitomagnetic apsidal period is identical to the Keplerian one can be intuitively justified since there is no net shift per orbit of the mean anomaly at epoch $\eta$. Indeed, from \rfr{kazzo}, one infers that $\eta$ is proportional to $t_\mathrm{p}$ through $\nk$.
Thus, since the latter stays constant because $a$ is not secularly affected by the gravitomagnetic field, the rate of change of the mean anomaly at epoch is proportional to the opposite of the pace of variation of the time of passage at pericentre according to
\eqi
\dert\eta t = -\nk\,\dert{t_\mathrm{p}} t.
\eqf
Should $\eta$ increase, the crossing of the pericentre would be anticipated with respect to the Keplerian  case since $t_\mathrm{p}$ would decrease, and vice versa. In this case, the variation of $\eta$ would result in an orbit--by--orbit advance or delay of the passages at the pericentre, which does not occur in the present case because, in fact, $\ang{d\eta/dt}^\mathrm{LT} = 0$.

\subsection{The Newtonian quadrupolar correction}\lb{ano_J2}
The $J_2$--affected anomalistic period can be calculated by means of \rfrs{ARJ2}{ANJ2} as explained in Section~\ref{ano_calc}. It turns out to be
\eqi
T_\mathrm{ano}^{J_2} = \Pb + \Delta T_\mathrm{ano}^{J_2},
\eqf
with
\begin{align}
\Delta T_\mathrm{ano}^{J_2} \lb{TanoJ2} & = \rp{3\uppi J_2 R_\mathrm{e}^2\ton{1+e\cos f_0}^3}{2\ton{1-e^2}^3\sqrt{\upmu a}}\ton{
-2\widehat{T}_1 + 3\widehat{T}_2 + 3\widehat{T}_3 \cos 2u_0 + 6 \widehat{T}_6\sin 2u_0}.
\end{align}

\begin{figure}
\centering
\includegraphics[width=\columnwidth]{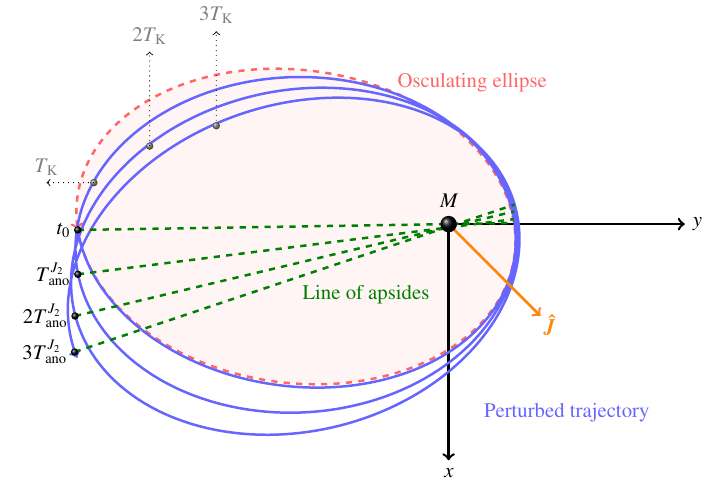}
\caption{Perturbed $J_2$ trajectory (continuous blue curve)  and its osculating
Keplerian ellipse (dashed red curve)  at the initial instant of time $t_0$ characterized by $e = 0.7,\,I = 30^\circ,\,\Omega = 45^\circ,\,\omega = 50^\circ,\,f_0 = 180^\circ$ as seen from the $z$--axis. The orientation of the spin axis $\boldsymbol{\hat{J}}$ of the central body is set by $\alpha_J=45^\circ,\,\delta_J=60^\circ$. In this example, $I$, $\Omega$, $\omega$ and $\eta$ undergo their known Newtonian precessions due to the quadrupole mass  moment $J_2$ of the primary \citep{2017EPJC...77..439I}; their magnitudes are suitably rescaled by enhancing them for a better visualization. The positions on the perturbed trajectory after one, two and three Keplerian periods $\Pb$ are marked  in gray. At each orbit, the passages at the drifting dashed green line of apsides occur always later than in the Keplerian case by the amount given by \rfr{TanoJ2}, which is positive  for the given values of the spin and orbital parameters.
}\label{fig_per_J2_ano}
\end{figure}

Figure \ref{fig_per_J2_ano}, obtained for generic values of the Keplerian  orbital elements, confirms the analytical result of \rfr{TanoJ2}; over three orbital revolutions, the test particle reaches always the precessing line of apsides after a time interval equal to $T_\mathrm{ano}^{J2}$ for each orbit. For the particular choice of the values of the primary's spin and orbital parameters, it turns out to be longer than $\Pb$, in agreement with \rfr{TanoJ2}.

Furthermore, Figure \ref{P_ano_J2} plots  the final part of the time series of the cosine $\boldsymbol{\hat{r}}\boldsymbol{\cdot}\boldsymbol{\hat{C}}$ of the angle between the position vector $\boldsymbol{r}$ and the Laplace--Runge--Lenz unit vector $\boldsymbol{\hat{C}}$ versus time $t$, in units of $\Pb$, for a numerically integrated fictitious test particle with and without \rfr{AJ2} starting in both cases from, say, the moving pericentre, i.e., for ${\boldsymbol{\hat{r}}}_0\boldsymbol{\cdot}{\boldsymbol{\hat{C}}}_0 = +1$. It can be seen that it comes back to the same position on the precessing line of apsides, i.e. it is $\boldsymbol{\hat{r}}\boldsymbol{\cdot}\boldsymbol{\hat{C}} = +1$ again, just after   $ T^{J_2}_\mathrm{ano} = \Pb + \Delta T^{J_2}_\mathrm{ano}$ differing from $\Pb$ by a (positive) amount in agreement with \rfr{TanoJ2} for the particular choice of the generic values of the spin and the orbital parameters adopted in the numerical integrations.
\begin{figure}
\centering
\includegraphics[width=\columnwidth]{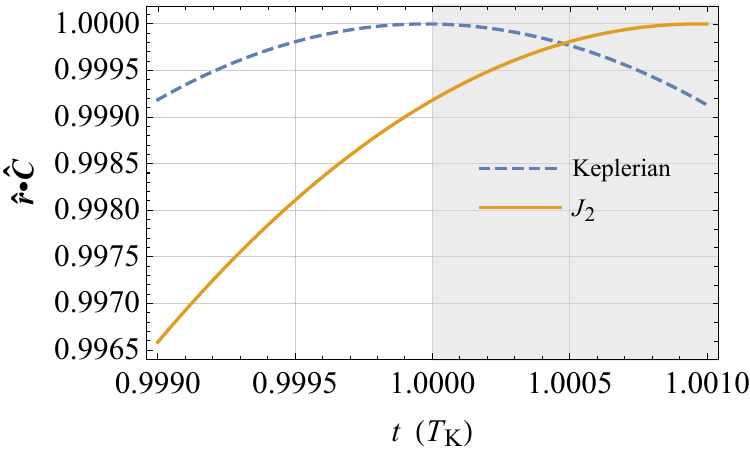}
\caption{Numerically produced time series of the cosine $\boldsymbol{\hat{r}}\boldsymbol{\cdot}\boldsymbol{\hat{C}}$ of the angle between the position vector $\boldsymbol{r}$ and the Laplace--Runge--Lenz vector $\boldsymbol{C}$  versus time $t$, in units of $\Pb$, obtained by integrating the equations of motion of a fictitious test particle with (continuous ocra yellow curve) and without (dashed azure curve) the $J_2$ acceleration of \rfr{AJ2} for an elliptical ($e=0.665$)  orbit arbitrarily oriented in space ($I = 40^\circ,\,\Omega = 45^\circ,\,\omega = 50^\circ$) starting from the periapsis ($f_0 = 0$), i.e., ${\boldsymbol{\hat{r}}}_0\boldsymbol\cdot{\boldsymbol{\hat{C}}}_0 = +1$; the semimajor axis is $a = 6 R_\mathrm{e}$. The physical parameters of the Earth are adopted, apart from the spin axis position set by $\alpha_J=45^\circ,\,\delta_J=60^\circ$. The $J_2$ acceleration is suitably rescaled in such a way that $\left|\Delta T^{J_2}_\mathrm{ano}\right|/\Pb=0.001$.   The time needed to come back to the initial position on the (moving) line of apsides, so that $\boldsymbol{\hat{r}}\boldsymbol{\cdot}\boldsymbol{\hat{C}} = +1$ again, is longer than in the Keplerian case by the amount $\Delta T^{J_2}_\mathrm{ano} = +0.001\Pb$, shown by the shaded area, in agreement with \rfr{TanoJ2}. }\lb{P_ano_J2}
\end{figure}
\section{The draconitic period}\lb{Dra_sect}
\subsection{General calculational scheme}\lb{dra_calc}
For a perturbed trajectory, the draconitic period $T_\mathrm{dra}$ is defined as the
time interval between two successive instants when the real position
of the test particle coincides with the ascending  position on the corresponding osculating ellipse.

It can be calculated as \citep{1977AN....298..107M,2016MNRAS.460.2445I}
\eqi
T_\mathrm{dra} = \Pb + \Delta T_\mathrm{dra} = \bigintss_0^{2\uppi}\ton{\dert{t}{u}}du,\lb{mioc77}
\eqf
where $dt/du$, when a pK perturbing acceleration $\boldsymbol{A}$ is present, can be obtained as follows.

From \rfr{arglat} and \rfr{dfdt}, one obtains \citep{Ocho59,1977AN....298..107M}
\eqi
\dert u t = \rp{\sqrt{\upmu p}}{r^2}\ton{1 - \rp{r^2 \cos I}{\sqrt{\upmu p}}\dert\Omega t}.
\eqf
Then, it is
\eqi
\dert t u \simeq \rp{r^2}{\sqrt{\upmu p}} + \rp{r^4 \cos I}{\upmu p}\dert\Omega t.
\eqf
Note that $d\Omega/dt$ is already expressed in terms of $u$, as per \rfr{dOdt}.
By using the nonsingular orbital elements\footnote{They are the components of the eccentricity vector  \citep{Taff85}, an alternative formulation of the Laplace--Runge--Lenz vector \citep{Gold80}. In the context of pulsar astronomy, they are also known as first and second Laplace--Lagrange parameters $\epsilon_1,\epsilon_2$ \citep{LorKra05}.} \citep{1977AN....298..107M,Monte2006}
\begin{align}
k &:=e\sin\omega,\acap
q &:=e\cos\omega,
\end{align}
\rfr{rKep} can be cast into the form
\eqi
r = \rp{p}{1 + q\cos u + k\sin u}
\eqf
in which $p,\,q,\,k$ are treated as independent variables.
By proceeding as in Section~\ref{ano_calc}, it can be finally obtained \citep{1977AN....298..107M,2016MNRAS.460.2445I}
\begin{align}
\Delta T_\mathrm{dra} \lb{bah} & = \bigint_0^{2\uppi}\grf{
\rp{3}{2}\sqrt{\rp{p}{\upmu}}\rp{\Delta p\ton{u}}{\ton{1 + q\cos u + k\sin u}^2} - 2\sqrt{\rp{p^3}{\upmu}}\rp{\cos u~\Delta q\ton{u} + \sin u~\Delta k\ton{u}}{\ton{1 + q\cos u + k\sin u}^3} + \rp{r^4 \cos I}{\upmu p}\dert\Omega t
}_\mathrm{K}du.
\end{align}
The first--order variations $\Delta p\ton{u}$, $\Delta q\ton{u}$ and $\Delta k\ton{u}$ entering \rfr{bah} are worked out by integrating  the following expressions \citep{1977AN....298..107M}
\begin{align}
\dert p u \lb{dpdu} & = \rp{2 r^3~A_{\tau}}{\upmu}, \acap
\dert q u \lb{dqdu} & = \rp{r^2\sin u~A_{r}}{\upmu} + \rp{r^2\qua{r~q + \ton{r + p}\cos u }~A_{\tau}}{\upmu} + \rp{\cot I~r^3~k\sin u~A_{h}}{\upmu p}, \acap
\dert k u \lb{dkdu} & = -\rp{r^2\cos u~A_{r}}{\upmu} + \rp{r^2\qua{r~k + \ton{r + p}\sin u }~A_{\tau}}{\upmu} - \rp{\cot I~r^3~q\sin u~A_{h}}{\upmu p}
\end{align}
from $u_0$ to $u$.

As far as the actual measurability of the draconitic period in some astronomical scenario of interest is concerned, it was shown \citep{Amelin66,Kassi66,Zhongo66} that it is possible to measure it for an artificial Earth's satellite\footnote{In their analyses, \citet{Amelin66,Kassi66,Zhongo66} used the Soviet satellite 1960 $\varepsilon$ 3.} as the ratio of the difference of the times of passages of the sub--satellite point through a chosen parallel for two following epochs to the number of satellite revolutions corresponding to this difference. The accuracy reached at that time should be of the order of $\simeq 10^{-4}$ s \citep{Kassi66}; it is not unlikely that it could be improved by orders
of magnitude with the most recent techniques currently available.
\subsection{The 1pN gravitoelectric correction}\lb{dra_GE}
The 1pN draconitic period can be calculated by means of \rfrs{Ar1pN}{Ah1pN} as explained in Section~\ref{dra_calc}. It turns out to be
\eqi
T_\mathrm{dra}^\mathrm{1pN} = \Pb + \Delta T_\mathrm{dra}^\mathrm{1pN},
\eqf
with
\begin{align}
\Delta T_\mathrm{dra}^\mathrm{1pN} \nonumber & = \rp{\uppi\sqrt{\upmu_\mathrm{b} a}}{4 c^2}\ton{
72 + e^2 \ton{84 - 76 \nu} + 4 e^4 \ton{6 - 7 \nu} - 16 \nu - 3 e \grf{\qua{8 \ton{-7 + 3 \nu} + e^2 \ton{-24 + 31 \nu}} \cos f_0  + \right.\right.\acap
\lb{draGENU} &\left.\left. + e \qua{4 \ton{-5 + 4 \nu} \cos 2f_0  + e \nu \cos 3f_0 }} - \rp{24\sqrt{1 - e^2}}{\ton{1 + e \cos\omega}^2}
}.
\end{align}
In the test particle limit $\nu\rightarrow 0$, \rfr{draGENU} reduces to
\begin{align}
\Delta T_\mathrm{dra}^\mathrm{1pN} \lb{TdraGE} &= \rp{3\uppi\sqrt{\upmu a}}{c^2}\qua{\rp{6 + 7 e^2 + 2 e^4 + 2 e \ton{7 + 3 e^2} \cos f_0 +
 5 e^2 \cos 2f_0}{\ton{1 - e^2}^2} -  \rp{2\sqrt{1-e^2}}{\ton{1 + e \cos\omega}^2}}.
\end{align}
It can be noted that \rfr{TdraGE} is always positive for all values of $e,\,f_0$ and $\omega$; thus, the node is reached later than in the Keplerian case.

\begin{figure}
\centering
\includegraphics[width=\columnwidth]{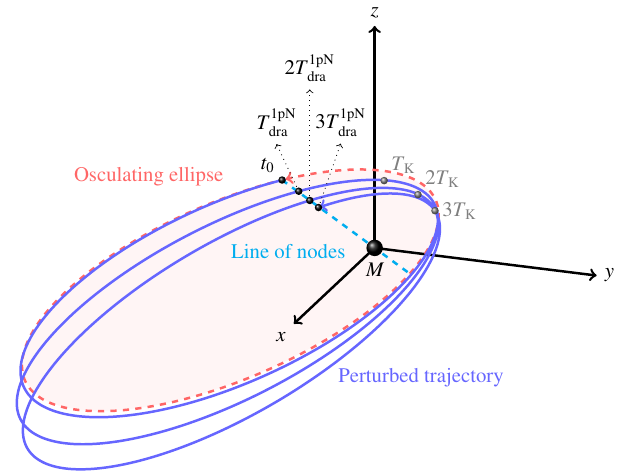}
\caption{Perturbed 1pN trajectory (continuous blue curve)  and its osculating
Keplerian ellipse (dashed red curve)  at the initial instant of time $t_0$ characterized by $e = 0.7,\,I = 30^\circ,\,\Omega = 45^\circ,\omega=50^\circ,\,f_0 = 180^\circ - \omega$. In this example, it is assumed that both $\omega$ and $\eta$ undergo the known 1pN gravitoelectric secular precessions due to the mass $M$ of the primary \citep{2017EPJC...77..439I}. For a better visualization of their effect, their sizes are suitably rescaled.
The positions on the perturbed trajectory after one, two and three Keplerian periods $\Pb$ are marked  in gray. At each orbit, the passages at the fixed dashed cyan line of nodes  occurs  always later than in the Keplerian case by the amount given by \rfr{TdraGE}, which is always positive.
}\label{fig_per_GE_dra}
\end{figure}

Figure \ref{fig_per_GE_dra}, obtained for generic values of the Keplerian  orbital elements, confirms the analytical result of \rfr{TdraGE}; over, say, three orbital revolutions, the test particle reaches always the fixed line of nodes after a time interval equal to $T_\mathrm{dra}^\mathrm{1pN}$. It is longer than $\Pb$, in agreement with \rfr{TdraGE}.
Furthermore, Figure \ref{P_dra_GE} plots  the final part of the time series of the cosine $\boldsymbol{\hat{r}}\boldsymbol{\cdot}\boldsymbol{\hat{l}}$ of the angle between the position vector $\boldsymbol{r}$ and the node unit vector $\boldsymbol{\hat{l}}$ versus time $t$, in units of $\Pb$, for a numerically integrated fictitious test particle with and without \rfr{A1pN} starting in both cases from, say, the fixed ascending node, i.e., for ${\boldsymbol{\hat{r}}}_0\boldsymbol{\cdot}{\boldsymbol{\hat{l}}}_0 = +1$. It can be seen that it comes back to the same position on the constant line of nodes, i.e. it is $\boldsymbol{\hat{r}}\boldsymbol{\cdot}\boldsymbol{\hat{l}} = +1$ again, just after   $T^\mathrm{1pN}_\mathrm{dra} = \Pb + \Delta T^\mathrm{1pN}_\mathrm{dra}$ differing from $\Pb$ by a (positive) amount, in agreement with \rfr{TdraGE}.
\begin{figure}
\centering
\includegraphics[width=\columnwidth]{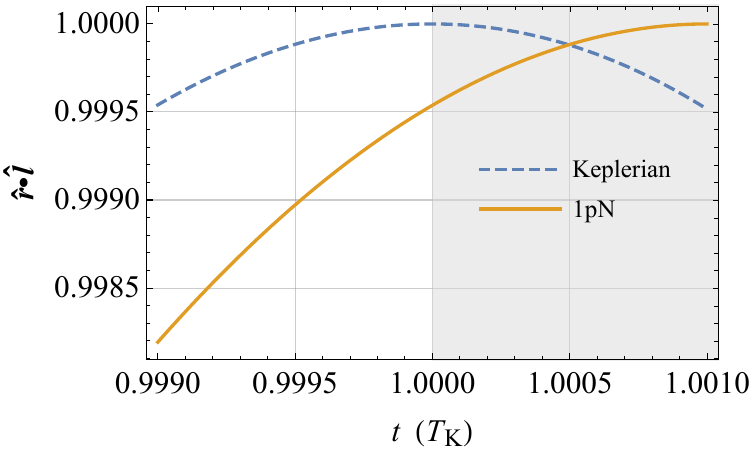}
\caption{Numerically produced time series of the cosine $\boldsymbol{\hat{r}}\boldsymbol{\cdot}\boldsymbol{\hat{l}}$ of the angle between the position vector $\boldsymbol{r}$ and the node unit vector $\boldsymbol{\hat{l}}$  versus time $t$, in units of $\Pb$, obtained by integrating the equations of motion of a fictitious test particle with (continuous ocra yellow curve) and without (dashed azure curve) the 1pN gravitoelectric acceleration of \rfr{A1pN} for an elliptical ($e=0.665$)  orbit arbitrarily oriented in space ($I = 40^\circ,\,\Omega = 45^\circ,\,\omega = 50^\circ$) starting from the ascending node $\ascnode$ ($f_0 = -\omega+360^\circ$), i.e., ${\boldsymbol{\hat{r}}}_0\boldsymbol\cdot{\boldsymbol{\hat{l}}}_0 = +1$; the semimajor axis is $a = 6 R_\mathrm{e}$. The physical parameters of the Earth are adopted. The  1pN acceleration is suitably rescaled in such a way that $\Delta T^\mathrm{1pN}_\mathrm{dra}/\Pb=0.001$. The time needed to come back to the initial position on the (fixed) line of nodes, so that $\boldsymbol{\hat{r}}\boldsymbol{\cdot}\boldsymbol{\hat{l}} = +1$ again, is longer than in the Keplerian case by the amount $\Delta T^\mathrm{1pN}_\mathrm{dra} = +0.001\Pb$, shown by the shaded area, in agreement with \rfr{TdraGE}. }\lb{P_dra_GE}
\end{figure}
\subsection{The 1pN gravitomagnetic Lense--Thirring correction}\lb{dra_LT}
The LT draconitic period can be calculated by means of \rfrs{ARLT}{ANLT} as explained in Section~\ref{dra_calc}. It turns out to be
\eqi
T_\mathrm{dra}^\mathrm{LT} = \Pb + \Delta T_\mathrm{dra}^\mathrm{LT},
\eqf
with
\eqi
\Delta T_\mathrm{dra}^\mathrm{LT} = \rp{4\uppi J\ton{2\texttt{Jh} + \texttt{Jm}\cot I}}{c^2 M\ton{1+e\cos\omega}^2}\lb{TdraLT}.
\eqf
The explicit form of the geometric coefficient in the numerator of \rfr{TdraLT} depending on the orientation in space of both the orbital plane and the primary's spin axis is
\begin{align}
2\texttt{Jh} + \texttt{Jm}\cot I \lb{koff} &= 3\cos I\sin\delta + \cos\delta\ton{\csc I - 3\sin I}\sin\ton{\alpha_J - \Omega}.
\end{align}
In general, it can be either positive and negative.
For a polar orbit, i.e. for $\Omega = \alpha_J$ and $I=90^\circ$, the gravitomagnetic correction vanishes, as per \rfr{koff}.
Instead, for an equatorial orbit arbitrarily oriented in space, it does not vanish amounting to
\eqi
\Delta T_\mathrm{dra}^\mathrm{LT} = \pm\rp{8\uppi J}{c^2 M\ton{1+e\cos\omega}^2}.\lb{TdraLT:equa}
\eqf
Furthermore, for circular orbits, \rfr{TdraLT:equa} reduces to
\eqi
\Delta T_\mathrm{dra}^\mathrm{LT} = \pm\rp{8\uppi J}{c^2 M}.\lb{TdraLT:equa:circ}
\eqf
If the orbital plane lies in the reference plane, i.e. for $I=0$, \rfr{TdraLT} loses its meaning, as it is expected since, in this case, the line of nodes is no longer defined.

\begin{figure}
\centering
\includegraphics[width=\columnwidth]{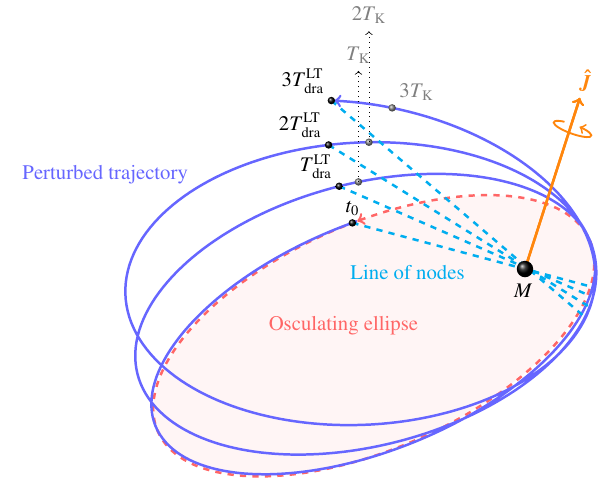}
\caption{Perturbed LT trajectory (continuous blue curve)  and its osculating
Keplerian ellipse (dashed red curve)  at the initial instant of time $t_0$ characterized by $e = 0.7,\,I = 30^\circ,\,\Omega = 72^\circ,\,\omega = 50^\circ,\,f_0 = 180^\circ-\omega$. The orientation of the spin axis $\boldsymbol{\hat{J}}$ of the central body is set by $\alpha_J=45^\circ,\,\delta_J=60^\circ$. In this example,  $I$, $\Omega$ and $\omega$ undergo their known LT precessions due to the spin angular momentum $\boldsymbol{J}$ of the primary \citep{2017EPJC...77..439I}; their magnitudes are suitably rescaled by enhancing them for a better visualisation.
The positions on the perturbed trajectory after one, two and three Keplerian periods $\Pb$ are marked   as well. At each orbit, the passages at the precessing dashed cyan line of nodes occur always later than in the Keplerian case by the amount given by \rfr{TdraLT}, which is positive for the given values of the spin and orbital parameters.
}\label{fig_per_LT_dra}
\end{figure}

Figure \ref{fig_per_LT_dra}, obtained for generic values of the Keplerian orbital parameters, confirms the analytical result of \rfr{TdraLT}; over three orbital revolutions, the test particle reaches always the precessing line of nodes after a time interval equal to $T_\mathrm{dra}^\mathrm{LT}$ after each orbit. For the particular choice of the values of the primary and orbital parameters, it turns out to be longer than $\Pb$, in agreement with \rfr{TdraLT}.

Furthermore, Figure \ref{P_dra_LT} plots  the final part of the time series of the cosine $\boldsymbol{\hat{r}}\boldsymbol{\cdot}\boldsymbol{\hat{l}}$ of the angle between the position vector $\boldsymbol{r}$ and the node unit vector $\boldsymbol{\hat{l}}$ versus time $t$, in units of $\Pb$, for a numerically integrated fictitious test particle with and without \rfr{ALT} starting in both cases from, say, the moving ascending
node, i.e., for ${\boldsymbol{\hat{r}}}_0\boldsymbol{\cdot}{\boldsymbol{\hat{l}}}_0 = +1$. It can be seen that it comes back to the same position on the precessing line of nodes, i.e. it is $\boldsymbol{\hat{r}}\boldsymbol{\cdot}\boldsymbol{\hat{l}} = +1$ again, just after   $ T^\mathrm{LT}_\mathrm{dra} = \Pb + \Delta T^\mathrm{LT}_\mathrm{dra}$ differing from $\Pb$ by a (positive) amount, in agreement with \rfr{TdraLT} for the particular choice of the generic values of the spin and the orbital parameters adopted in the numerical integrations.
\begin{figure}
\centering
\includegraphics[width=\columnwidth]{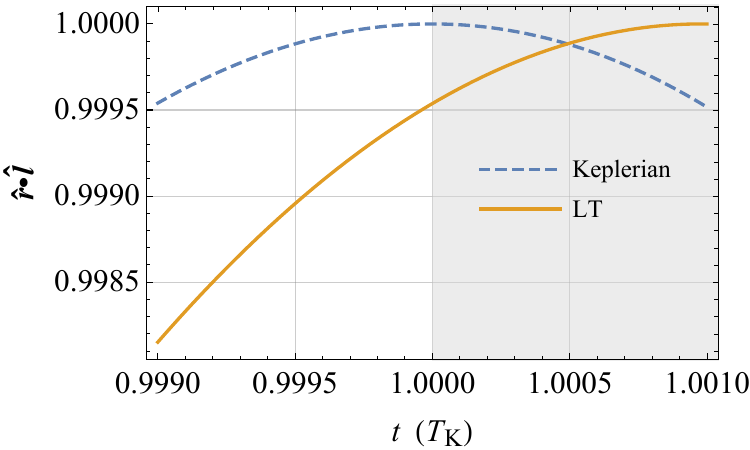}
\caption{Numerically produced time series of the cosine $\boldsymbol{\hat{r}}\boldsymbol{\cdot}\boldsymbol{\hat{l}}$ of the angle between the position vector $\boldsymbol{r}$ and the node unit vector $\boldsymbol{\hat{l}}$  versus time $t$, in units of $\Pb$, obtained by integrating the equations of motion of a fictitious test particle with (continuous ocra yellow curve) and without (dashed azure curve) the LT acceleration of \rfr{ALT} for an elliptical ($e=0.665$)  orbit arbitrarily oriented in space ($I = 40^\circ,\,\Omega = 45^\circ,\,\omega = 50^\circ$) starting from the ascending node $\ascnode$ ($f_0 = -\omega+360^\circ$), i.e., ${\boldsymbol{\hat{r}}}_0\boldsymbol\cdot{\boldsymbol{\hat{l}}}_0 = +1$; the semimajor axis is $a = 6 R_\mathrm{e}$. The physical parameters of the Earth are adopted, apart from the spin axis position set by $\alpha_J=45^\circ,\,\delta_J=60^\circ$. The LT acceleration is suitably rescaled in such a way that $\left|\Delta T^\mathrm{LT}_\mathrm{dra}\right|/\Pb=0.001$. The time needed to come back to the initial position on the (moving) line of nodes, so that $\boldsymbol{\hat{r}}\boldsymbol{\cdot}\boldsymbol{\hat{l}} = +1$ again, is longer than in the Keplerian case by the amount $\Delta T^\mathrm{LT}_\mathrm{dra} = +0.001\Pb$, shown by the shaded area, in agreement with \rfr{TdraLT}. }\lb{P_dra_LT}
\end{figure}
\subsection{The Newtonian quadrupolar correction}\lb{dra_J2}
The $J_2$--affected draconitic period can be calculated by means of \rfrs{ARJ2}{ANJ2} as explained in Section~\ref{dra_calc}. It turns out to be
\eqi
T_\mathrm{dra}^{J_2} = \Pb + \Delta T_\mathrm{dra}^{J_2},
\eqf
with
\begin{align}
\Delta T_\mathrm{dra}^{J_2} \lb{TdraJ2} & = \rp{3\uppi J_2 R_\mathrm{e}^2}{2\sqrt{\upmu a\ton{1-e^2}}}\qua{\rp{1}{\ton{1 + e\cos\omega}^2}\ton{-2\widehat{T}_1 + 3\widehat{T}_2 - 2\widehat{T}_5 \cot I} + \rp{\ton{1 + e\cos f_0}^3}{\ton{1-e^2}^{5/2}}\ton{-2\widehat{T}_1 + 3\widehat{T}_2 + 3\widehat{T}_3 \cos 2u_0  + 6\widehat{T}_6\sin 2u_0}}.
\end{align}
It can be noted that \rfr{TdraJ2} is not defined for $I\rightarrow 0$ because of the term
\begin{align}
\widehat{T}_5\cot I \nonumber &= \cot I\qua{\sin I \sin\delta_J + \cos I \cos\delta_J \sin\ton{\alpha_J - \Omega}}\qua{\cos I \sin\delta_J - \cos\delta_J \sin I \sin\ton{\alpha_J - \Omega}},
\end{align}
as it is expected since, in this case, the line of nodes is no longer defined.

\begin{figure}
\includegraphics[width=\columnwidth]{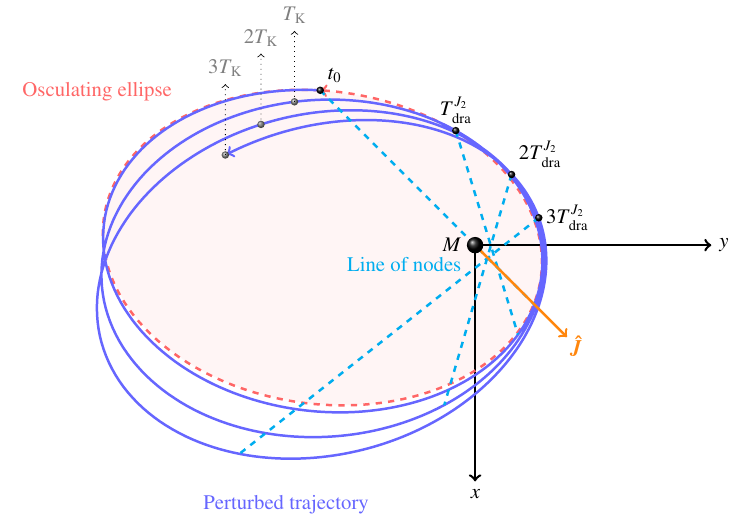}
\caption{Perturbed $J_2$ trajectory (continuous blue curve)  and its osculating
Keplerian ellipse (dashed red curve)  at the initial instant of time $t_0$ characterized by $e = 0.7,\,I = 30^\circ,\,\Omega = 45^\circ,\,\omega = 50^\circ,\,f_0 = 180^\circ-\omega$ as seen from the $z$--axis. The orientation of the spin axis $\boldsymbol{\hat{J}}$ of the central body is set by $\alpha_J=45^\circ,\,\delta_J=60^\circ$. In this example,  $I$, $\Omega$, $\omega$ and $\eta$ undergo their known Newtonian shifts due to the quadrupole mass  moment $J_2$ of the primary \citep{2017EPJC...77..439I};  their magnitudes are suitably rescaled  for better visualizing their effect.
The positions on the perturbed trajectory after one, two and three Keplerian periods $\Pb$ are marked in gray. At each orbit, the passages at the precessing dashed cyan line of nodes  occur always earlier than in the Keplerian case by the amount given by \rfr{TdraJ2}, which is negative for the given values of the spin and orbital parameters.
}\label{fig_per_J2_dra}
\end{figure}

Figure \ref{fig_per_J2_dra}, obtained for generic values of the Keplerian  orbital elements, confirms the analytical result of \rfr{TdraJ2}; over three orbital revolutions, the test particle reaches always the precessing line of nodes after a time interval equal to $T_\mathrm{dra}^{J_2}$ after each orbit. For the particular choice of the values of the primary's spin and orbital parameters, it is shorter than $\Pb$, in agreement with \rfr{TdraJ2}.

Furthermore, Figure \ref{P_dra_J2} plots  the final part of the time series of the cosine $\boldsymbol{\hat{r}}\boldsymbol{\cdot}\boldsymbol{\hat{l}}$ of the angle between the position vector $\boldsymbol{r}$ and the node unit vector $\boldsymbol{\hat{l}}$ versus time $t$, in units of $\Pb$, for a numerically integrated fictitious test particle with and without \rfr{AJ2} starting in both cases from, say, the moving ascending node, i.e., for ${\boldsymbol{\hat{r}}}_0\boldsymbol{\cdot}{\boldsymbol{\hat{l}}}_0 = +1$. It can be seen that it comes back to the same position on the precessing line of nodes, i.e. it is $\boldsymbol{\hat{r}}\boldsymbol{\cdot}\boldsymbol{\hat{l}} = +1$ again, just after   $ T^{J_2}_\mathrm{dra} = \Pb + \Delta T^{J_2}_\mathrm{dra}$ differing from $\Pb$ by a (positive) amount in agreement with \rfr{TdraJ2} for the particular choice of the generic values of the spin and the orbital parameters adopted in the numerical integrations.
\begin{figure}
\centering
\includegraphics[width=\columnwidth]{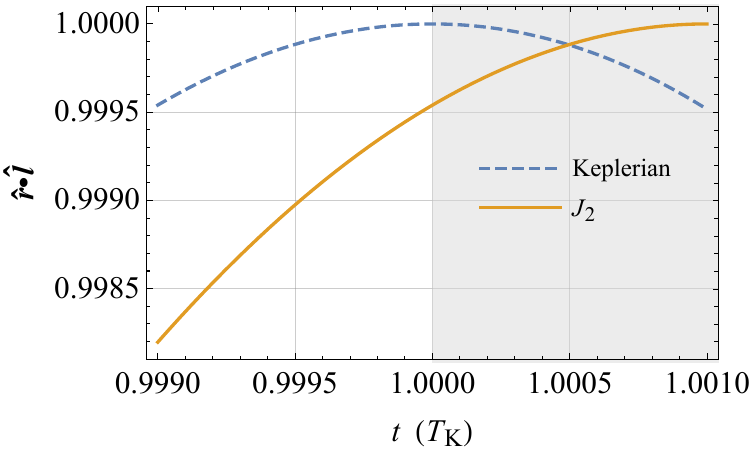}
\caption{Numerically produced time series of the cosine $\boldsymbol{\hat{r}}\boldsymbol{\cdot}\boldsymbol{\hat{l}}$ of the angle between the position vector $\boldsymbol{r}$ and the node unit vector $\boldsymbol{\hat{l}}$  versus time $t$, in units of $\Pb$, obtained by integrating the equations of motion of a fictitious test particle with (continuous ocra yellow curve) and without (dashed azure curve) the $J_2$ acceleration of \rfr{AJ2} for an elliptical ($e=0.665$)  orbit arbitrarily oriented in space ($I = 40^\circ,\,\Omega = 45^\circ,\,\omega = 50^\circ$) starting from the ascending node $\ascnode$ ($f_0 = -\omega+360^\circ$), i.e., ${\boldsymbol{\hat{r}}}_0\boldsymbol\cdot{\boldsymbol{\hat{l}}}_0 = +1$; the semimajor axis is $a = 6 R_\mathrm{e}$. The physical parameters of the Earth are adopted, apart from the spin axis position set by $\alpha_J=45^\circ,\,\delta_J=60^\circ$. The $J_2$ acceleration is suitably rescaled in such a way that $\left|\Delta T^{J_2}_\mathrm{dra}\right|/\Pb=0.001$. The time needed to come back to the initial position on the (moving) line of nodes, so that $\boldsymbol{\hat{r}}\boldsymbol{\cdot}\boldsymbol{\hat{l}} = +1$ again, is longer than in the Keplerian case by the amount $\Delta T^{J_2}_\mathrm{dra} = +0.001\Pb$, shown by the shaded area, in agreement with \rfr{TdraJ2}. }\lb{P_dra_J2}
\end{figure}
\section{The sidereal period}\lb{Sid_sect}
\subsection{General calculational scheme}\lb{sid_calc}
In general, both the line of nodes and the line of apsides do vary over time because of one or more pK accelerations. Thus, it may be useful to look at a characteristic orbital timescale involving the crossing of some fixed reference direction in space; the sidereal period $T_\mathrm{sid}$, defined as the time interval between two successive instants when the real position of the test particle lies on
a given reference direction, plays well such a role.

For an orbit arbitrarily inclined, the sidereal period can be calculated as
\eqi
T_\mathrm{sid} = \Pb + \Delta T_\mathrm{sid} = \bigintss_0^{2\uppi}\ton{\dert t \phi}d\phi,\lb{Tsd}
\eqf
where $\phi\ton{t}$ is the azimuthal angle reckoned  from the reference $x$ axis in the fundamental plane; when the latter is assumed to be coincident with, say, the Earth's equatorial plane at some reference epoch, $\phi\ton{t}$ is the right ascension $\alpha\ton{t}$ of the celestial body of interest.
From
\begin{align}
x\ton{t} \lb{Ics}& = r\ton{t}\qua{\cos\Omega\cos u\ton{t} - \cos I\sin\Omega\sin u\ton{t}},\ \\ \nonumber \\
y\ton{t} \lb{Ypsilon}& = r\ton{t}\qua{\sin\Omega\cos u\ton{t} + \cos I\cos\Omega\sin u\ton{t}},
\end{align}
one obtains $\phi\ton{t}$ as
\eqi
\phi\ton{t} =\arctan\qua{\rp{y\ton{t}}{x\ton{t}}};\lb{azimphi}
\eqf
it is a function of the generally varying $I\ton{t}$, $\Omega\ton{t}$, and of $u\ton{t}$, i.e. $\phi\ton{t}=\phi\ton{I\ton{t},\Omega\ton{t},u\ton{t}}$.
Since the ongoing calculation is to the first order in the pK acceleration, the differential $d\phi$ in \rfr{Tsd} can be written as
\eqi
d\phi \simeq \ton{\derp\phi u} du.
\eqf
The integrand of \rfr{Tsd} can be obtained as
\begin{align}
\dert t\phi  &=\rp{1}{\dert\phi t}=\rp{1}{\derp\phi I\dert I t+ \derp\phi\Omega \dert\Omega t + \derp\phi u\dert u t} = \rp{1}{\dert u t\derp\phi u\qua{1 + \derp u\phi\ton{\derp\phi I\dert I u + \derp\phi\Omega\dert\Omega u}}}.
\end{align}
Thus, to the first order in the pK acceleration, the integral of \rfr{Tsd} can be approximated as
\begin{align}
T_\mathrm{sid}\lb{megaTsd} &\simeq\bigintss_0^{2\uppi}\dert t u\qua{1 - \derp u\phi\ton{\derp\phi I\dert I u+\derp\phi\Omega\dert\Omega u}}du = \bigintss_0^{2\uppi}\ton{\dert t u}du -\bigintss_0^{2\uppi}\rp{1}{\derp\phi u}\ton{\derp\phi I\dert I u + \derp\phi\Omega\dert\Omega u}\ton{\dert t u} du.
\end{align}
The first term in \rfr{megaTsd} is nothing but the draconitic period, and can be calculated to the order $\mathcal{O}\ton{A}$ as outlined in Section~\ref{dra_calc}. The second term in \rfr{megaTsd}
is a correction to the former
\eqi
\Delta T_\mathrm{sid\,II} := -\bigintss_0^{2\uppi}\rp{1}{\derp\phi u}\ton{\derp\phi I\dert I u + \derp\phi\Omega\dert\Omega u}\ton{\dert t u}_\mathrm{K}du.\lb{bigs}
\eqf
taking into account the fact that, in general, the orbital plane is displaced by the pK acceleration; indeed, the rates of $I$ and $\Omega$ enter it. In \rfr{bigs}, the suffix K appended to $dt/du$ implies that it has to be calculated onto the unperturbed Keplerian  ellipse in order to keep the calculation to the first order in $A$.

If the orbital plane coincides with the fundamental one, the previously outlined calculational strategy may lead to analytical expressions for $T_\mathrm{sid}$ which, for some pK accelerations, are singular in $I=0$. In that cases, the sidereal period can be straightforwardly calculated by means of the true longitude
\eqi
l:=\varpi + f,\lb{trulo}
\eqf
where
\eqi
\varpi:=\Omega + \omega
\eqf
is the longitude of pericentre\footnote{It is a dogleg angle \textcolor{black}{\citep{2000ssd..book.....M,Shev17}} since $\Omega$ and $\omega$ lie generally in different planes.},
as \citep{2016MNRAS.460.2445I}
\eqi
T_\mathrm{sid} = \Pb + \Delta T_\mathrm{sid} = \bigintss_0^{2\uppi}\ton{\dert t l}dl,\lb{ior16}
\eqf
in close analogy with Section~\ref{ano_calc} and Section~\ref{dra_calc}.
It should be recalled that $l$ is generally a dogleg angle since $\Omega$ and $u$ are located in different planes; it is the true longitude of the test particle actually moving along its real orbit only if $I=0$.
When a pK perturbing acceleration $\boldsymbol{A}$ enters the equations of motion, $dt/dl$ can be obtained in the following way.

From \rfr{trulo} and \rfr{dtdf2}, it is
\eqi
\dert l t = \rp{\sqrt{\upmu p}}{r^2}\qua{1 + \rp{2 r^2 \sin^2\ton{I/2}}{\sqrt{\upmu p}}\dert\Omega t}.
\eqf
Then, it can be written
\eqi
\dert t l \simeq \rp{r^2}{\sqrt{\upmu p}} - \rp{2 r^4 \sin^2\ton{I/2}}{\upmu p}\dert\Omega t.
\eqf
The sine of the argument of latitude $u$ entering \rfr{dOdt} for $d\Omega/dt$ can be written in terms of $l$ as $\sin\ton{l - \Omega}$.
By introducing the nonsingular
equinoctial elements \citep{1972CeMec...5..303B}
\begin{align}
Q &:=e\cos\varpi,\acap
K &:=e\sin\varpi,
\end{align}
\rfr{rKep} can be rewritten as
\eqi
r = \rp{p}{1 + Q\cos l + K\sin l}
\eqf
in which $p,\,Q,\,K$ are to be treated as as independent variables.
By proceeding as in Section~\ref{ano_calc} and Section~\ref{dra_calc}, one obtains \citep{2016MNRAS.460.2445I}
\begin{align}
\Delta T_\mathrm{sid} \lb{bih} & = \bigint_0^{2\uppi}\grf{
\rp{3}{2}\sqrt{\rp{p}{\upmu}}\rp{\Delta p\ton{l}}{\ton{1 + Q\cos l + K\sin l}^2} - 2\sqrt{\rp{p^3}{\upmu}}\rp{\cos l~\Delta Q\ton{l} + \sin l~\Delta K\ton{l}}{\ton{1 + Q\cos l + K\sin l}^3} - \rp{2 r^4 \sin^2\ton{I/2}}{\upmu p}\dert\Omega t
}_\mathrm{K}dl.
\end{align}
The first--order variations $\Delta p\ton{l}$, $\Delta Q\ton{l}$ and $\Delta K\ton{l}$ entering \rfr{bih} can be worked out by integrating  the following expressions \citep{2016MNRAS.460.2445I}
\begin{align}
\dert p l \lb{dpdl} & = \rp{2 r^3~A_{\tau}}{\upmu}, \acap
\dert{Q} l \lb{dQdl} & = \rp{r^2\sin l~A_{r}}{\upmu} + \rp{r^2\qua{r~Q + \ton{r + p}\cos l}~A_{\tau}}{\upmu} -
\rp{\tan\ton{I/2} r^3~K \sin\ton{l-\Omega}~A_{h}}{\upmu p}, \acap
\dert{K} l \lb{dKdl} & = -\rp{r^2\cos l~A_{r}}{\upmu} + \rp{r^2\qua{r~K + \ton{r + p}\sin l}~A_{\tau}}{\upmu} +
\rp{\tan\ton{I/2} r^3~Q\sin\ton{l - \Omega}~A_{h}}{\upmu p}
\end{align}
from $l_0$ to $l$.
If the orbital plane is aligned with the fundamental one, \rfr{bih} and \rfrs{dQdl}{dKdl} have to be calculated with $I=0$.

It is generally expected that if the orbital plane stays constant in space, i.e, if neither the nodes, when defined, nor the orbit's projection onto the fundamental plane change over time, the sidereal period coincides with the draconitic one since the line of nodes is a fixed direction in space.
\subsection{The 1pN gravitoelectric correction}\lb{sid_GE}
As shown in Section~\ref{sid_calc}, the sidereal period for a generic perturbed orbit is the sum of the draconitic period, calculated as explained in Section~\ref{dra_calc}, and the term given by \rfr{bigs}.
For \rfr{A1pN}, \rfr{bigs} turns out to be
\eqi
\Delta T^\mathrm{1pN}_\mathrm{sid\,II} = 0.
\eqf
Thus, in this case, the sidereal period  coincides with the draconitic  one.

This is shown in Figure \ref{P_sid_GE}. It plots the final part of the time series of the cosine of the angle $\phi$, normalized to its initial value $\cos \phi_0$, versus time $t$, in units of $\Pb$, for a numerically integrated fictitious test particle with and without \rfr{A1pN} starting from the same generic initial position. It can be seen that it comes back to the same position on the fixed direction chosen in the reference plane, i.e. it is $\cos\phi/\cos\phi_0 = +1$ again, just after   $T^\mathrm{1pN}_\mathrm{sid} = T^\mathrm{1pN}_\mathrm{dra}$  differing from $\Pb$ by a positive amount, in agreement with \rfr{TdraGE}.
\begin{figure}
\centering
\includegraphics[width=\columnwidth]{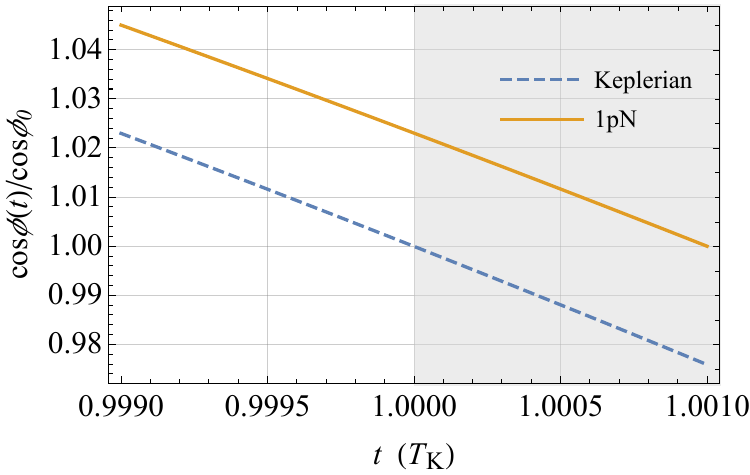}
\caption{Numerically produced time series of the cosine $\cos\phi\ton{t}$ of the azimuthal angle $\phi\ton{t}$ normalized to its initial value $\cos\phi_0$  versus time $t$, in units of $\Pb$, obtained by integrating the equations of motion of a fictitious test particle with (continuous ocra yellow curve) and without (dashed azure curve) the 1pN gravitoelectric acceleration of \rfr{A1pN} for an elliptical ($e=0.665$)  orbit arbitrarily oriented in space ($I = 40^\circ,\,\Omega = 45^\circ,\,\omega = 50^\circ$) starting from, say, the ascending node $\ascnode$ ($f_0 = -\omega+360^\circ$); the semimajor axis is $a = 6 R_\mathrm{e}$. The physical parameters of the Earth are adopted. The 1pN acceleration is suitably rescaled in such a way that $\Delta T^\mathrm{1pN}_\mathrm{sid}/\Pb=0.001$. The time  needed to $\cos\phi\ton{t}$ to assume again its initial value $\cos\phi_0$ is longer than in the Keplerian case by the amount $\Delta T^\mathrm{1pN}_\mathrm{sid} = +0.001\Pb$, shown by the shaded area, in agreement with the sum of \rfr{TdraGE}. }\lb{P_sid_GE}
\end{figure}
\subsection{The 1pN gravitomagnetic Lense--Thirring correction}\lb{sid_LT}
As shown in Section~\ref{sid_calc}, the sidereal period for a generic perturbed orbit is the sum of the draconitic period, calculated as explained in Section~\ref{dra_calc}, and the term given by \rfr{bigs}.
For \rfr{ALT}, \rfr{bigs} yields
\begin{align}
\Delta T^\mathrm{LT}_\mathrm{sid\,II} \lb{TsidLTaux} & = \rp{4\uppi \boldsymbol{J}\cot I}{c^2 M e^2\sqrt{1-e^2}}\boldsymbol{\cdot}\grf{
\boldsymbol{\hat{m}}\qua{-e^2 + 2 \ton{2 - e^2 - 2 \sqrt{1 - e^2}} \cos 2\omega} + 2\boldsymbol{\hat{l}} \ton{-2 + e^2 + 2 \sqrt{1 - e^2}}\sin 2\omega
}.
\end{align}
In the equatorial case, the orbital plane stays constant in space, \rfr{TsidLTaux} vanishes, and the sidereal period coincides with the draconitic one, as it is expected since neither the line of nodes nor the orbit's projection onto the reference plane change.
By taking the sum of \rfr{TdraLT} and \rfr{TsidLTaux}, the full expression of the gravitomagnetic correction of the sidereal period $\Delta T_\mathrm{sid}^\mathrm{LT}$ is obtained. It can be noted that, for a generic eccentric orbit, $\Delta T_\mathrm{sid}^\mathrm{LT}$ is not defined if the orbital plane lies in the fundamental one. Nonetheless, for $e=0$, it reduces to
\eqi
\Delta T_\mathrm{sid}^\mathrm{LT} = \rp{8\uppi J}{c^2 M}\qua{\cos I \sin\delta_J -\cos\delta_J \sin I \sin\ton{\alpha_J - \Omega}},\lb{nosing}
\eqf
which is not singular in $I=0$.
By using the true longitude $l$ in the case $I=0$, it turns out
\eqi
\Delta T_\mathrm{sid}^\mathrm{LT} = \rp{8\uppi J \sin\delta_J}{c^2 M\ton{1+e\cos\varpi}^2}.
\eqf
In the limit $e\rightarrow 0$, it reduces to
\eqi
\Delta T_\mathrm{sid}^\mathrm{LT} = \rp{8\uppi J\sin\delta_J}{c^2 M},\lb{trulo}
\eqf
which agrees with \rfr{nosing} calculated with $I=0$. In turn, if $\delta_J = \pm 90^\circ$, corresponding to the case of an equatorial orbit whose orbital plane coincides with the reference plane, \rfr{trulo} becomes
\eqi
\Delta T_\mathrm{sid}^\mathrm{LT} = \pm\rp{8\uppi J}{c^2 M},\lb{trulo2}
\eqf
in agreement with \rfr{TdraLT:equa:circ}.

\begin{figure}
\includegraphics[width=\columnwidth]{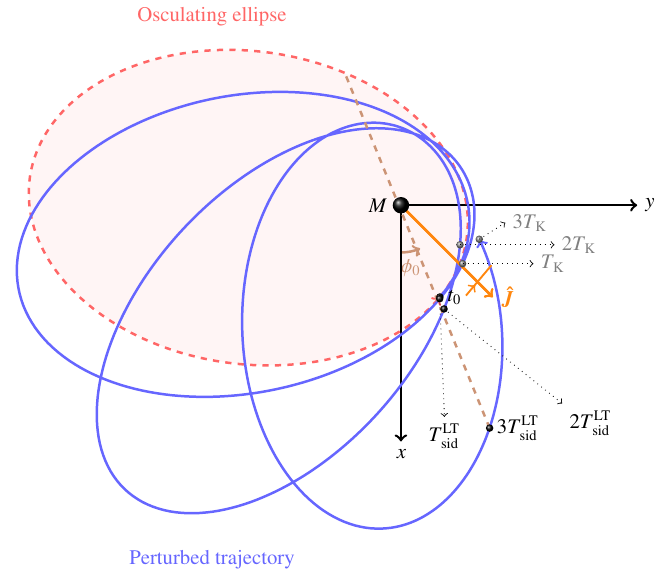}
\caption{Projections of the perturbed LT trajectory (continuous blue curve)  and of its osculating
Keplerian ellipse (dashed red curve)  in the reference  plane $\grf{x,\,y}$ at the initial instant of time $t_0$ characterized by the generic initial conditions $e = 0.7,\,I = 30^\circ,\,\Omega = 45^\circ,\,\omega = 50^\circ,\,f_0 = 285^\circ$. The orientation of the spin axis $\boldsymbol{\hat{J}}$ of the central body, whose projection in the fundamental plane is depicted as well, is set by $\alpha_J=45^\circ,\,\delta_J=60^\circ$. In this example, $I$, $\Omega$, and $\omega$ undergo their known LT shifts due to the spin angular momentum $\boldsymbol{J}$ of the primary \citep{2017EPJC...77..439I};  their sizes are suitably rescaled  for better visualizing their effect.
The positions on the perturbed trajectory after one, two and three Keplerian periods $\Pb$ are marked   as well. At each orbit, the passages at the generic fixed dashed brown line characterized by $\phi_0$ occur always earlier than in the Keplerian case by the amount given by the sum of \rfr{TdraLT} and \rfr{TsidLTaux}. It is so because, for the given values of the spin and orbital parameters, $\Delta T_\mathrm{dra}^\mathrm{LT} + \Delta T_\mathrm{sid\,II}^\mathrm{LT}<0$, as per \rfr{TdraLT} and \rfr{TsidLTaux}.
}\label{fig_per_LT_sid}
\end{figure}

Figure \ref{fig_per_LT_sid} confirms the analytical results of \rfr{TdraLT} and \rfr{TsidLTaux}. Indeed, over three orbital revolutions, the projection   of a generic LT perturbed orbit in the fundamental plane $\grf{x,\,y}$  crosses  a fixed direction in the latter set by a certain value $\phi_0$ always after a time interval equal to $T_\mathrm{sid}^\mathrm{LT} = T_\mathrm{dra}^\mathrm{LT} + \Delta T_\mathrm{sid\,II}^\mathrm{LT}$ for each orbit. With the particular choice of the primary's spin and the orbital parameters used in the picture, $T_\mathrm{sid}^\mathrm{LT}$ turns out to be shorter than $\Pb$, in agreement with \rfr{TdraLT} and \rfr{TsidLTaux}.

Furthermore, Figure \ref{P_sid_LT} plots the final part of the time series of the cosine of the angle $\phi$, normalized to its initial value $\cos \phi_0$, versus time $t$, in units of $\Pb$, for a numerically integrated fictitious test particle with and without \rfr{ALT} starting from the same generic initial position. It can be seen that it comes back to the same position on the fixed direction chosen in the reference plane, i.e. it is $\cos\phi/\cos\phi_0 = +1$ again, just after   $T^\mathrm{LT}_\mathrm{sid} = T^\mathrm{LT}_\mathrm{dra} + \Delta T^\mathrm{LT}_\mathrm{sid\,II}$ differing from $\Pb$ by a (positive) amount in agreement with \rfr{TdraLT} and \rfr{TsidLTaux} for the particular choice of the generic values of the spin and the orbital parameters adopted in the numerical integrations.
\begin{figure}
\centering
\includegraphics[width=\columnwidth]{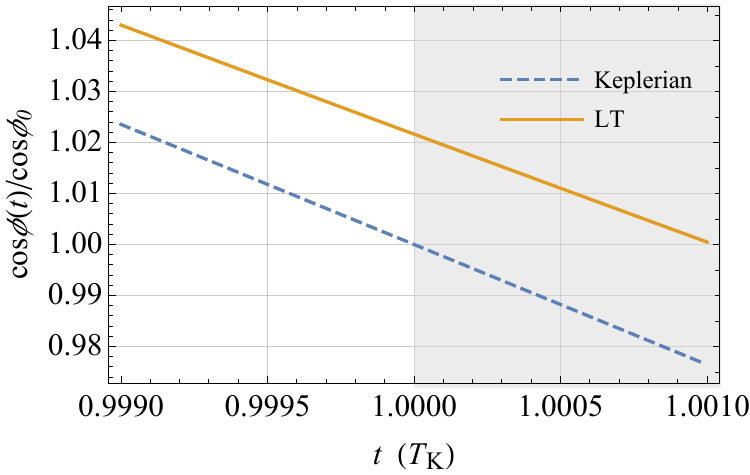}
\caption{Numerically produced time series of the cosine $\cos\phi\ton{t}$ of the azimuthal angle $\phi\ton{t}$ normalized to its initial value $\cos\phi_0$  versus time $t$, in units of $\Pb$, obtained by integrating the equations of motion of a fictitious test particle with (continuous ocra yellow curve) and without (dashed azure curve) the LT acceleration of \rfr{ALT} for an elliptical ($e=0.665$)  orbit arbitrarily oriented in space ($I = 40^\circ,\,\Omega = 45^\circ,\,\omega = 310^\circ$) starting from, say, $f_0 = 50^\circ$; the semimajor axis is $a = 6 R_\mathrm{e}$. The physical parameters of the Earth are adopted, apart from the spin axis position set by $\alpha_J=45^\circ,\,\delta_J=60^\circ$. The LT acceleration is suitably rescaled in such a way that $\left|\Delta T^\mathrm{LT}_\mathrm{sid}\right|/\Pb=0.001$. The time  needed to $\cos\phi\ton{t}$ to assume again its initial value $\cos\phi_0$ is longer than in the Keplerian case by the amount $\Delta T^\mathrm{LT}_\mathrm{sid} = +0.001\Pb$, shown by the shaded area, in agreement with the sum of \rfr{TdraLT} and \rfr{TsidLTaux}. }\lb{P_sid_LT}
\end{figure}

\subsection{The Newtonian quadrupolar correction}\lb{sid_J2}
As shown in Section~\ref{sid_calc}, the sidereal period for a generic perturbed orbit is the sum of the draconitic period, calculated as explained in Section~\ref{dra_calc}, and the term given by \rfr{bigs}.
For \rfr{AJ2}, \rfr{bigs} turns out to be
\begin{align}
\Delta T^{J_2}_\mathrm{sid\,II} \lb{TsidJ2aux} & = \rp{3\uppi J_2 R_\mathrm{e}^2\cot I}{e^2\sqrt{\upmu a\ton{1 - e^2}}}\grf{
\widehat{T}_5\qua{e^2 + 2 \ton{-2 + e^2 + 2 \sqrt{1 - e^2}} \cos 2\omega} - 2\widehat{T}_4 \ton{-2 + e^2 + 2 \sqrt{1 - e^2}}\sin 2\omega
}.
\end{align}
For equatorial orbits, \rfr{TsidJ2aux} vanishes, and the sidereal period reduces to the draconitic one.
The oblateness of the sidereal period $\Delta T_\mathrm{sid}^{J_2}$ can be obtained by summing \rfr{TdraJ2} and \rfr{TsidJ2aux}; for an elliptic orbit, it turns out to be singular in $I=0$. Instead, in the limit $e\rightarrow 0$, it reduces to
\begin{align}
\Delta T_\mathrm{sid}^{J_2} \nonumber &= \rp{3\uppi J_2 R_\mathrm{e}^2}{2\sqrt{\upmu a}}\grf{
-4 + 6 \cos^2\delta_J \cos^2\ton{\alpha_J-\Omega}  +  6 \cos\delta_J \cos\ton{\alpha_J-\Omega} \sin 2u_0 \qua{\sin I  \sin\delta_J  + \cos I  \cos\delta_J \sin\ton{\alpha_J-\Omega}} +  \right.\acap
\nonumber &\left. + 6 \qua{\sin I  \sin\delta_J + \cos I  \cos\delta_J \sin\ton{\alpha_J-\Omega}}^2 + 3 \cos 2u_0 \qua{\cos\delta_J \cos\ton{\alpha_J-\Omega} - \sin I  \sin\delta_J - \cos I  \cos\delta_J \sin\ton{\alpha_J-\Omega}}\times\right.\acap
\lb{atroce} &\left.\times\qua{\cos\delta_J \cos\ton{\alpha_J-\Omega} + \sin I  \sin\delta_J + \cos I  \cos\delta_J \sin\ton{\alpha_J-\Omega}}
},
\end{align}
which is defined also for that value of the inclination. In such a case, using the true longitude $l$ yields
\begin{align}
\Delta T_\mathrm{sid}^{J_2} \nonumber & = -\rp{3\uppi J_2 R_\mathrm{e}^2}{4\ton{1-e^2}^2\sqrt{\upmu a}}\qua{\rp{\ton{-2+3\cos^2\delta_J}}{\ton{1+e\cos\varpi}^2}
\qua{2 + e^2 - 2\ton{1 - e^2}^{3/2}  + 4 e \cos\varpi + e^2 \cos 2\varpi} + \right.\acap
\nonumber &\left. + \rp{1}{2\ton{1-e^2}}\ton{\ton{4 + e^2}\ton{1 - 3 \cos 2\delta_J} - e\qua{-1 + 6 \cos^2\ton{l_0 - \alpha_J} \cos 2\delta_J}\qua{3 e \cos\ton{2l_0-2\varpi} + 6 \cos\ton{l_0 - \varpi}  + \right.\right.\right.\acap
\lb{aiutoo}&\left.\left.\left. + 2 e^2 \cos^3\ton{l_0-\varpi}} - 3 \cos\ton{2l_0  - 2\alpha_J}\grf{\ton{2 + 3 e^2} \cos 2\delta_J  + 2\qua{1 + e \cos\ton{l_0-\varpi}}^3}
}
}.
\end{align}
In the limit $e\rightarrow 0$, \rfr{aiutoo} agrees with \rfr{atroce} calculated for $I=0$.

\begin{figure}
\includegraphics[width=\columnwidth]{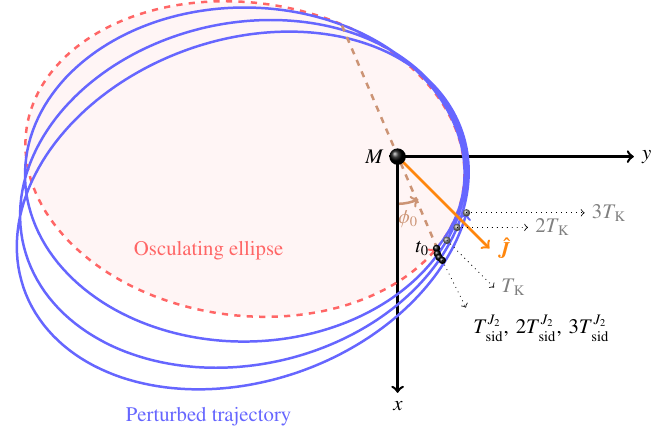}
\caption{Projections of the perturbed $J_2$ trajectory (continuous blue curve)  and of its osculating
Keplerian ellipse (dashed red curve)  in the reference plane $\grf{x,\,y}$ at the initial instant of time $t_0$ characterized by the generic initial conditions $e = 0.7,\,I = 30^\circ,\,\Omega = 45^\circ,\,\omega = 50^\circ,\,f_0 = 285^\circ$. The orientation of the spin axis $\boldsymbol{\hat{J}}$ of the central body, whose projection in the fundamental plane is depicted as well, is set by $\alpha_J=45^\circ,\,\delta_J=60^\circ$. In this example, $I$, $\Omega$, $\omega$ and $\eta$ undergo the known Newtonian shifts  due to the quadrupole mass  moment $J_2$ of the primary \citep{2017EPJC...77..439I};  their magnitudes are suitably rescaled  for better visualizing their effect.
The positions on the perturbed trajectory after one, two and three Keplerian periods $\Pb$ are marked   as well. At each orbit, the passages at the generic fixed dashed brown line characterized by $\phi_0$ occur always earlier than in the Keplerian case by the amount given by the sum of \rfr{TdraJ2} and \rfr{TsidJ2aux}. It is so because, for the given values of the spin and orbital parameters, $\Delta T_\mathrm{dra}^{J_2} + \Delta T_\mathrm{sid\,II}^{J_2}<0$, as per \rfr{TdraJ2} and \rfr{TsidJ2aux}.
}\label{fig_per_J2_sid}
\end{figure}

Figure \ref{fig_per_J2_sid} confirms the analytical results of \rfr{TdraJ2} and \rfr{TsidJ2aux}. Indeed, over three orbital revolutions, the projection   of a generic $J_2$--perturbed orbit in the fundamental plane $\grf{x,\,y}$  crosses  a fixed direction in the latter set by a certain value $\phi_0$ always after a time interval equal to $T_\mathrm{sid}^{J_2} = T_\mathrm{dra}^{J_2} + \Delta T_\mathrm{sid\,II}^{J_2}$ after each orbit. For the particular choice of the primary's spin and the orbital parameters used in the picture, $T_\mathrm{sid}^{J_2}$ turns out to be shorter than $\Pb$, in agreement with \rfr{TdraJ2} and \rfr{TsidJ2aux}.

Furthermore, Figure \ref{P_sid_J2} plots the final part of the time series of the cosine of the angle $\phi$, normalized to its initial value $\cos \phi_0$, versus time $t$, in units of $\Pb$, for a numerically integrated fictitious test particle with and without \rfr{AJ2} starting from the same generic initial position. It can be seen that it comes back to the same position on the fixed direction chosen in the reference plane, i.e. it is $\cos\phi/\cos\phi_0 = +1$ again, just after   $T^{J_2}_\mathrm{sid} = T^{J_2}_\mathrm{dra} + \Delta T^{J_2}_\mathrm{sid\,II}$ differing from $\Pb$ by a (positive) amount in agreement with \rfr{TdraJ2} and \rfr{TsidJ2aux} for the particular choice of the generic values of the spin and the orbital parameters adopted in the numerical integrations.
\begin{figure}
\centering
\includegraphics[width=\columnwidth]{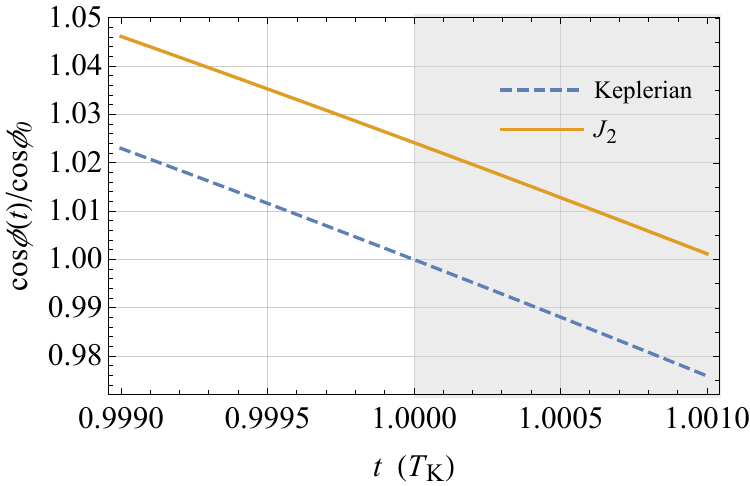}
\caption[$J_2$ sidereal period: $\cos\phi\ton{t}/\cos\phi_0$]{Plot of the numerically produced time series of the cosine $\cos\phi\ton{t}$ of the azimuthal angle $\phi\ton{t}$ normalized to its initial value $\cos\phi_0$  versus time $t$, in units of $\Pb$, obtained by integrating the equations of motion of a fictitious test particle with (continuous ocra yellow curve) and without (dashed azure curve) the $J_2$ acceleration of \rfr{AJ2} for an elliptical ($e=0.665$)  orbit arbitrarily oriented in space ($I = 40^\circ,\,\Omega = 45^\circ,\,\omega = 50^\circ$) starting from, say, the ascending node $\ascnode$ ($f_0 = -\omega+360^\circ$); the semimajor axis is $a = 6 R_\mathrm{e}$. The physical parameters of the Earth are adopted, apart from the spin axis position set by $\alpha_J=45^\circ,\,\delta_J=60^\circ$. The $J_2$ acceleration is suitably rescaled in such a way that $\left|\Delta T^{J_2}_\mathrm{sid}\right|/\Pb=0.001$. The time  needed to $\cos\phi\ton{t}$ to assume again its initial value $\cos\phi_0$ is longer than in the Keplerian case by the amount $\Delta T^{J_2}_\mathrm{sid} = +0.001\Pb$, shown by the shaded area, in agreement with the sum of \rfr{TdraJ2} and \rfr{TsidJ2aux}. }\lb{P_sid_J2}
\end{figure}
\section{Some numerical evaluations}\lb{num_sect}
The accuracy in measuring the orbital period of several transiting exoplanets \citep{2010trex.book.....H} is nowadays at the
\eqi
\sigma_T\simeq 10^{-7}-10^{-8}\,\mathrm{d}\simeq 9\times 10^{-3}-10^{-4}\,\mathrm{s},
\eqf
level\footnote{See \url{https://exoplanet.eu/home/} and \url{https://exoplanetarchive.ipac.caltech.edu/}.}

\textcolor{black}{Even better accuracies are available for other classes of objects; suffice it to say that} the period of WD1032 + 011 b, an inflated brown dwarf in an old eclipsing binary with a white dwarf \textcolor{black}{with}
\eqi
\textcolor{black}{\nu=0.11,}
\eqf
is known with an uncertainty as small as \citep{2020MNRAS.497.3571C}
\eqi
\sigma_T\simeq 4.5\times 10^{-10}\,\mathrm{d} = 3.8\times 10^{-5}\,\mathrm{s}.\lb{sigmaT}
\eqf

The relevant orbital and physical parameters of WD1032 + 011 b are \citep{2020MNRAS.497.3571C}
\begin{align}
M_\mathrm{p}/M_\odot \lb{par1} & = 0.0665\pm 0.0061\,\acap
M_\mathrm{s}/M_\odot & = 0.4502\pm 0.0500\,\acap
a/R_\odot \lb{par3} & = 0.6854\pm 0.0244,
\end{align}
where p and s designate the planet and the star, respectively; $M_\odot$ and $R_\odot$ are the mass and the radius of the Sun, respectively.

By assuming that the measured orbital period is the sidereal\footnote{A circular orbit is assumed for  WD1032 + 011 b \citep{2020MNRAS.497.3571C}.} one, \rfr{draGENU}, calculated with \rfrs{par1}{par3}, yields
\eqi
\Delta T_\mathrm{dra}^\mathrm{1pN} = \Delta T_\mathrm{sid}^\mathrm{1pN} = 0.07\pm 0.004\,\mathrm{s}. \lb{theo}
\eqf
\Rfr{sigmaT} shows that, in principle, the 1pN gravitoelectric correction to the Keplerian orbital period, given by \rfr{theo}, falls within the measurability regime. \textcolor{black}{On the other hand, any excessive optimism should be tempered since the  Keplerian term should be finally subtracted from the measured period in order to extract the 1pN component. This implies that the values of the parameters entering the former one should be known to a sufficiently high level of accuracy, which is not yet the case. Indeed, from the errors in \rfrs{par1}{par3}, one can calculate that the resulting uncertainty in the Keplerian period is as large as}
\eqi
\textcolor{black}{\sigma_{T_\mathrm{K}} \simeq 572 s.}
\eqf
\textcolor{black}{An evaluation of the other pK corrections is not possible since they depend on $\boldsymbol{J}$, $J_2$ and the mutual spin--orbit orientation about which no information is provided by \citet{2020MNRAS.497.3571C}.}

\textcolor{black}{The importance of the issue of the mismodeling in the Keplerian period can be clearly understood in the case of the double pulsar PSR J0737--3039 \citep{2003Natur.426..531B,2004Sci...303.1153L}, characterized by}
\eqi
\textcolor{black}{\nu = 0.2497,}
\eqf
\textcolor{black}{and whose (anomalistic\footnote{\textcolor{black}{A. Possenti, personal communication, May 2024.}}) orbital period  is measured with an accuracy as good as \citep{2006Sci...314...97K}}
\eqi
\textcolor{black}{\sigma_{T}= 5\times 10^{-11}\,\mathrm{d} \simeq 4.32\times 10^{-6}\,\mathrm{s}.}\lb{doubleP}
\eqf
\textcolor{black}{In principle, the exquisite accuracy of \rfr{doubleP} would allow for an accurate determination of the 1pN gravitoelectric correction to the anomalistic period of PSR J0737--3039. Indeed, according to \rfr{anoGENU}, it nominally ranges from}
\eqi
\textcolor{black}{0.27\,\mathrm{s}\lesssim \Delta T_\mathrm{ano}^\mathrm{1pN}\lesssim 0.40\,\mathrm{s}},
\eqf
\textcolor{black}{depending on $f_0$. Unfortunately, the uncertainty in the Keplerian period, calculated by propagating the errors in the relevant parameters entering it \citep{2006Sci...314...97K},  turns out to be\footnote{\textcolor{black}{Incidentally, the difference between the measured orbital period and the Keplerian one is not statistically significative since it can be calculated to be as small as $1.9\,\mathrm{s}$.}}}
\eqi
\textcolor{black}{\sigma_{T_\mathrm{K}}\simeq 9\,\mathrm{s}.}
\eqf

\textcolor{black}{Should it be possible to measure for the same system at least two of the three characteristic orbital periods independently, their common Keplerian component would be automatically canceled by forming their difference.}
\section{Summary and conclusions}\lb{fine}
It was shown that  post--Keplerian accelerations perturbing a two--body gravitationally bound system, like those arising from the oblateness of the central body to the Newtonian level and from the post--Newtonian \textcolor{black}{gravitoelectromagnetic} mass and spin--dependent components of its external gravitational potential, breaks the degeneracy between the otherwise coincident anomalistic, draconitic and sidereal orbital periods.

The resulting corrections to the Keplerian orbital period are generally different for the aforementioned characteristic timescales. The sidereal period still coincides with the draconitic one when the 1pN gravitoelectric acceleration is taken into account, being both different from the anomalistic period. The 1pN gravitomagnetic LT acceleration leaves the anomalistic period unaffected with respect to the Keplerian case, while it modifies differently the draconitic and the sidereal periods. Finally, the oblateness of the central body alters all three orbital periods in different ways from each other. In general, all the non--vanishing corrections to the Keplerian orbital period depend on the true anomaly at epoch.

The resulting analytical expressions are completely general since they hold for arbitrary values of the orbital eccentricity and inclination. Furthermore, they are valid also for generic orientations of the primary's symmetry axis in space.

For the transiting \textcolor{black}{brown dwarf} WD1032 + 011 b \textcolor{black}{($\nu=0.11$)}, the predicted 1pN gravitoelectric correction to the (sidereal) orbital period amounts to 0.07 s, while the current uncertainty in measuring it is as small as $\simeq 10^{-5}$ s\textcolor{black}{, and the mismodeling in the Keplerian part is as large as 572 s}. \textcolor{black}{For the double pulsar PSR J0737--3039 ($\nu=0.2497$), the 1pN gravitoelectric correction to the anomalistic period, which is the measured one for this astrophysical system,  is as large as a few tenths of a second; despite the experimental accuracy in measuring the apsidal period is $\simeq 10^{-6}\,\mathrm{s}$, the present--day uncertainty in the calculated value of the Keplerian component is still too large, amounting to about 9 s. If it were possible to measure at least two of the three characteristic orbital timescales independently for the same system, the difference between them would allow the Keplerian term to be canceled a priori.}
\section*{Data availability}
No new data were generated or analysed in support of this research.
\section*{Conflict of interest statement}
I declare no conflicts of interest.
\bibliography{Megabib}{}
\end{document}